\documentclass{article}
\usepackage{iftex}
\RequirePDFTeX
\usepackage{arxiv}
\usepackage[T1]{fontenc}
\usepackage[authoryear,round]{natbib}

\usepackage{amsmath,amsfonts,bm}

\def\eqref#1{equation~\ref{#1}}

\def\1{\bm{1}}

\DeclareMathAlphabet{\mathsfit}{\encodingdefault}{\sfdefault}{m}{sl}
\SetMathAlphabet{\mathsfit}{bold}{\encodingdefault}{\sfdefault}{bx}{n}

\usepackage{xspace}

\newcommand{\llama}{\textsc{Llama-3.1-8B-Instruct}\xspace}
\newcommand{\qwen}{\textsc{Qwen-4B-Instruct}\xspace}
\newcommand{\gptmini}{\textsc{GPT-4o-Mini}\xspace}
\newcommand{\gptnano}{\textsc{GPT-4.1-Nano}\xspace}
\newcommand{\gptosstwenty}{\textsc{GPT-OSS-20B}\xspace}
\newcommand{\gptossonetwenty}{\textsc{GPT-OSS-120B}\xspace}

\usepackage{microtype}
\usepackage{tabularx}
\usepackage{enumitem}
\usepackage{pifont}
\usepackage{hyperref}
\usepackage{url}
\usepackage{booktabs}
\usepackage{graphicx}
\usepackage{subcaption}
\usepackage{tcolorbox}
\usepackage{listings}
\usepackage{xcolor}
\usepackage{wrapfig}
\usepackage{needspace}
\usepackage{placeins}
\usepackage{amssymb}
\usepackage[nameinlink,capitalize]{cleveref}

\tcbuselibrary{listings}

\lstdefinestyle{jsonstyle}{
    backgroundcolor=\color{gray!10},
    basicstyle=\ttfamily\small,
    breakatwhitespace=false,
    breaklines=true,
    keepspaces=true,
    showspaces=false,
    showstringspaces=false,
    tabsize=2,
    frame=single,
    rulecolor=\color{gray!50}
}

\graphicspath{{./}}

\usepackage{xurl}
\usepackage{etoolbox}
\apptocmd{\thebibliography}{\raggedright}{}{}

\makeatletter
\patchcmd{\@maketitle}
  {\vskip 0.4in \@minus 0.1in \center{\@date} \vskip 0.2in}
  {\vskip 0.12in}
  {}{\PackageError{paper}{Could not compact the arXiv title block}{}}
\patchcmd{\@maketitle}
  {\textsc{\undertitle}\\ \vskip 0.1in}
  {}
  {}{\PackageError{paper}{Could not remove the empty preprint line}{}}
\patchcmd{\@maketitle}
  {\vskip 0.2in}
  {\vskip 0.1in}
  {}{\PackageError{paper}{Could not compact the title's top spacing}{}}
\patchcmd{\@bottomtitlebar}
  {\vskip 0.29in}
  {\vskip 0.17in}
  {}{\PackageError{paper}{Could not compact the title's bottom spacing}{}}
\patchcmd{\@toptitlebar}
  {\vskip 0.25in}
  {\vskip 0.15in}
  {}{\PackageError{paper}{Could not compact the title rule spacing}{}}
\patchcmd{\@maketitle}
  {\bf\rule{\z@}{24\p@}\@author}
  {\bf\rule{\z@}{16\p@}\@author}
  {}{\PackageError{paper}{Could not compact the author spacing}{}}
\makeatother

\definecolor{darkblue}{rgb}{0,0,0.5}
\hypersetup{
    colorlinks=true,
    citecolor=darkblue,
    linkcolor=darkblue,
    urlcolor=darkblue
}

\usepackage[normalem]{ulem}
\definecolor{tbkcolor}{RGB}{210,100,0}

\newcommand\tbklinks{\hypersetup{linkcolor=tbkcolor,citecolor=tbkcolor,urlcolor=tbkcolor}}

\newcommand\tbkbox[2]{\mbox{#1{#2}}}
\DeclareRobustCommand\taylorremoved[1]{{\color{tbkcolor}\tbklinks%
  \let\tbkCref\Cref\let\tbkcref\cref\let\tbkref\ref\let\tbkeqref\eqref
  \let\tbkcitep\citep\let\tbkcitet\citet\let\tbkcite\cite
  \def\Cref##1##{\tbkbox{\tbkCref##1}}\def\cref##1##{\tbkbox{\tbkcref##1}}%
  \def\ref##1##{\tbkbox{\tbkref##1}}\def\eqref##1##{\tbkbox{\tbkeqref##1}}%
  \def\citep##1##{\tbkbox{\tbkcitep##1}}\def\citet##1##{\tbkbox{\tbkcitet##1}}\def\cite##1##{\tbkbox{\tbkcite##1}}%
  \sout{#1}}}

\title{Black-Box Forensics \\
for Conversational LLM Agents}

\author{
Isadora White, Yasaman Jafari, Taylor Berg-Kirkpatrick \\
University of California, San Diego
}

\date{}
\renewcommand{\headeright}{}
\renewcommand{\undertitle}{}
\renewcommand{\shorttitle}{Black-Box Forensics for Conversational LLM Agents}
\hypersetup{
    pdftitle={Black-Box Forensics for Conversational LLM Agents},
    pdfauthor={Isadora White, Yasaman Jafari, Taylor Berg-Kirkpatrick}
}
\begin{document}

\maketitle
\addtocontents{toc}{\protect\setcounter{tocdepth}{-1}}

\begin{abstract}

LLM-powered scams are proliferating, API providers silently change the models behind their endpoints, and jailbreaks are increasingly model-specific.
Each of these calls for black-box forensics of conversational LLM agents: learning what powers an agent from nothing but ordinary, non-adversarial conversation with it.
We study two capabilities.
\emph{Attribution} identifies which of a known set of base models or system prompts powers an agent, and we treat it as classification: a classifier trained on stylistic and n-gram features identifies the base model with 98\% accuracy from a few turns of conversation.
System prompts can be attributed the same way, but only by collecting labeled conversations for every candidate prompt, and the prompts deployed in the wild are too many and change too often for that.
\emph{Fingerprinting} instead decides whether two conversations come from the same agent, and so needs no training data from the agents it is applied to.
Our cross-encoder scores a pair of conversations zero-shot: on system prompts never seen in training, it reaches an AUC of 0.77 from a single pair of conversations, and 0.94 when 50 conversations per agent are aggregated.
It remains useful when every reply is paraphrased by another model, and it gives positive results on real-world case studies.
Conversational agents can therefore be linked and identified from ordinary conversation, with no access to their weights or their hidden instructions.
\end{abstract}


\section{Introduction}

\begingroup
\setlength{\intextsep}{6pt}
\begin{figure*}[!h]
    \centering
    \includegraphics[width=0.99\textwidth,trim=0 96bp 0 10bp,clip]{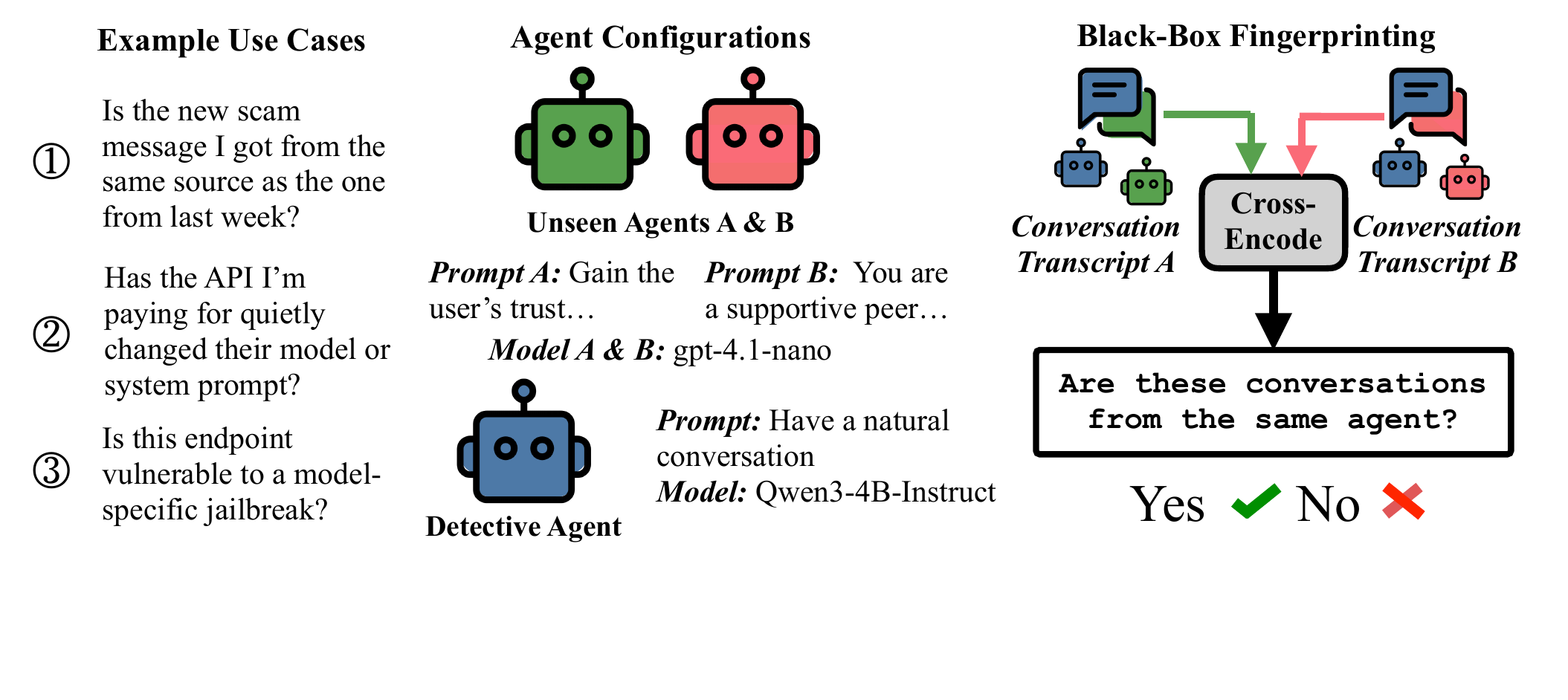}
    \caption{
    \normalsize{
    Each target agent is defined by a hidden system prompt and base model (center), observable only through conversation with our detective agent. Forensic questions (left) map onto our two capabilities: \textit{black-box attribution} (not depicted) identifies the base model behind an endpoint, while \textit{black-box fingerprinting} (right) determines whether two conversation transcripts originate from the same agent---even one never seen during training---by cross-encoding the pair. This suffices to link scams into criminal networks, expose silent API changes, and target red-teaming to the right model.}}
    \label{fig:zero_shot}
\end{figure*}
\FloatBarrier
\endgroup

As LLM adoption grows, so does abuse. AI-assisted fraud and social-engineering campaigns now operate at scale \citep{ic3_2024_report,ftc_2024_report}, bots spread misinformation unchecked across the internet, and the models behind API endpoints are swapped without disclosure. Investigators of this abuse observe nothing but conversational behavior, with no access to model weights or hidden instructions. This paper develops black-box forensics for conversational LLM agents: techniques that recover identifying information about an agent's base model and system prompt through conversation. We study two complementary capabilities: \textit{attribution}, identifying which of a known set of base models or system prompts powers an endpoint, which we treat as classification, and \textit{fingerprinting}, detecting whether two endpoints share the same---possibly never-before-seen---system prompt.

Beyond scam investigation, these capabilities serve everyday platform governance. API users depend on stable model behavior, yet providers may silently update backend checkpoints, safety layers, or system prompts, and such unannounced revisions can materially shift downstream behavior \citep{chen2023chatgpt, gao2024model}. Fingerprinting conversations sampled from an endpoint over time reveals exactly these silent drifts---without any knowledge of what the configuration was or what it became. Security teams face the complementary problem: jailbreaks transfer only partially across models and prompt templates \citep{zou2023universal,wei2023jailbroken}, so red-teaming must be tailored to a specific vulnerability profile, and attribution tells defenders which model they are actually facing.

A practical forensic tool must also be covert. Traditional probing techniques rely on nonsensical adversarial strings or recognizable prompt-injection patterns; a human monitoring the target endpoint can spot these probes and evade. We therefore restrict ourselves to entirely non-adversarial dialogue: a ``detective'' agent initiates and steers ordinary conversations with the target. Beyond covertness, this active elicitation paradigm allows us fine-grained control over topic and conversational flow, letting us isolate the target's behavioral signature from semantic noise.

We first study \textit{black-box attribution}: mapping a conversation to a base model or system prompt drawn from a fixed candidate set. Adapting ideas from static human-vs-LLM detection and authorship attribution \citep{uchendu2020authorship, venkatraman2024gpt, guggilla2025ai}, we attribute conversations to six base models with 98\% accuracy using stylistic and n-gram features, distinguishing even models from the same family of different sizes.
The same machinery extends to system prompts, but accuracy is correlated with the semantic similarity of the candidates.

Attribution, however, presumes a known set of candidates, each with labeled conversations to train on---and system prompts in the wild are unbounded and ever-changing. Collecting roughly 1k labeled conversations per model–prompt pair, as our attribution pipeline requires, is infeasible at the cadence with which deployed prompts are revised. We therefore introduce \textit{black-box fingerprinting}: determining whether two conversational threads originate from the same system prompt on the same base model, even when that configuration has never been seen during training. Because fingerprinting requires no examples from the target configuration, it scales to agents never seen in training: investigators can cluster distinct scam campaigns, and auditors can detect silent drift, without ever training on the new configuration. In our work, we evaluate our fingerprinting system directly on conversations with airline chatbots and real-world honeypot deployments.

Prior work has largely pursued three adjacent directions: (i) detecting whether text is human- or machine-generated \citep{mitchell2023detectgpt,kirchenbauer2023watermark}, (ii) extracting hidden instructions from proprietary systems via prompt-stealing attacks \citep{perez2022ignore, levin2025has}, and (iii) injecting identifiable signatures into models via instruction fine-tuning \citep{xu2024instructional}, probes \citep{bhardwaj2025invisible}, or watermarking \citep{ye2026securing, kirchenbauer2023watermark}. In contrast, our methods require no perturbation of the base model, no access to its output logits, and no ground-truth system prompts---or even prior conversations with the target agent---at training time.

Our key contributions and findings are as follows:

\begin{enumerate}[leftmargin=1.9em, labelsep=0.5em, topsep=0.5em, itemsep=0.35em, parsep=0pt, label=\textcolor{darkblue}{\ding{\numexpr181+\value{enumi}\relax}}]
    \item \textbf{Attribution of base models and system prompts.} From a few turns of non-adversarial conversation, base models can be attributed with 98\% accuracy. (\Cref{sec:attribution-techniques} and \Cref{sec: results}.)
    \item \textbf{Fingerprinting of unseen system prompts.} Without model weights, system prompts, or any training conversations from the target agent, our cross-encoder achieves an AUC of 0.768 and an F1 of 0.703 on entirely unseen system prompts. (\Cref{sec:fingerprinting-techniques} and \Cref{tab:zero_shot}.)
    \item \textbf{Fingerprinting scales with evidence.} Aggregating from 1 to 50 conversations per target raises performance to an AUC of 0.94 and an F1 of 0.79 (\Cref{fig:roc_auc_curves})
    \item \textbf{In-the-wild validation and robustness.} Our method holds up on real deployments, reaching 0.927 AUC on anonymized airline chatbots and 0.87 AUC on honeypot conversations with scammers, and is robust to shifts in punctuation, sampling, and choice of detective agent (AUC drop $<0.03$). (\Cref{sec:in-the-wild-dataset}, \Cref{tab:isolation_batch_metrics}, and \Cref{tab:real_world_results}.)
\end{enumerate}


\FloatBarrier
\section{Related Work}
\newcommand{\cmark}{\textcolor{green!60!black}{\checkmark}}
\newcommand{\xmark}{\textcolor{red!70!black}{\times}}

\begin{table*}[t]
\caption{\normalsize{\textbf{Comparison of our framework against existing auditing and attribution methods.}
We compare across key operational axes. Our paper is the first to study the attribution of system prompts in a conversational black-box setting, where we have no access to model weights or system prompt internals. 
\textbf{System prompts} refers to attributing or fingerprinting system prompts to outputs. \textbf{Attribution} refers to our classification-based attribution pipeline. \textbf{Active querying} refers to the ability of our detective agent to steer the conversation to control the topic, and \textbf{conversational} refers to our conversational setting. We refer to our methods as \textbf{black-box} as they do not rely on model internals or ground-truth system prompts to function. }}
\centering
\scriptsize
\setlength{\tabcolsep}{3pt}
\begin{tabular}{lccccc}
\toprule
\textbf{Method} & \textbf{System Prompts} & \textbf{Attribution} & \textbf{Active Querying} & \textbf{Conversational} & \textbf{Black-Box}\\
\midrule
LLMmap \citep{pasquini2025llmmap} & $\xmark$ & \cmark & \cmark & $\xmark$ & \cmark \\
Model Equality Testing \citep{gao2024model} & $\xmark$ & $\xmark$ & \cmark & $\xmark$ & \cmark \\
Anubis \citep{canonne2025zero} & $\xmark$ & $\xmark$ & $\xmark$ & $\xmark$ & $\xmark$ \\
\citep{guggilla2025ai} & $\xmark$ & \cmark & $\xmark$ & $\xmark$ & \cmark \\
GPT-Who \citep{venkatraman2024gpt} & $\xmark$ & \cmark & $\xmark$ & $\xmark$ & \cmark \\
DetectGPT \citep{mitchell2023detectgpt} & $\xmark$ & $\xmark$ & $\xmark$ & $\xmark$ & $\xmark$ \\
Prompt Stealing \citep{perez2022ignore} & \cmark & $\xmark$ & \cmark & $\xmark$ & \cmark \\
TRAP \citep{gubri2024trap} & $\xmark$ & \cmark & \cmark & $\xmark$ & \cmark \\
Black-Box Attribution \citep{bai2025esf} & $\xmark$ & \cmark & $\xmark$ & $\xmark$ & \cmark \\
\midrule
\textbf{Ours} & \cmark & \cmark & \cmark & \cmark & \cmark \\
\bottomrule
\end{tabular}

\label{tab:method_comparison}
\end{table*}

\textbf{LLM output detection.}
Recent work on distinguishing LLM from human text has achieved over 99\% accuracy and low false positive rates, such as the pangram AI detector \citep{emi2024technical}, DetectGPT \citep{mitchell2023detectgpt}, GLTR \citep{prajapati2024detection}, and GPT-Who \citep{venkatraman2024gpt}. 
Notably, some works, such as \citet{joshi2024hullmi}, show that using traditional ML techniques can perform similarly to more modern NLP techniques. 
Other works focus on the attribution of a specific model from generations \citep{guggilla2025ai, venkatraman2024gpt}. 
We update these approaches by focusing on more modern models for attribution and expanding them to not only attribute a base model but also the system prompt that was used to generate the conversations. 

\textbf{Prompt extraction and prompt-injection attacks.} While detecting differences in system prompts may not have been studied, prompt extraction of proprietary system prompts, especially those containing private data, has been studied extensively \citep{perez2022ignore,greshake2023more,das2025system, levin2025has, agarwal-etal-2024-prompt, wang2024raccoon}. A common attack for this is a sandwich attack, where a harmful query is placed between harmless queries \citep{upadhayay2024sandwich}, or leveraging the sycophancy effect \citep{agarwal-etal-2024-prompt}. Our approach is similar, but avoids \emph{directly querying for the information} in any discernible fashion - if the agent we are trying to fingerprint suspects our intentions, they might be able to evade.

\textbf{LLM Model Family and Prompt Attribution} \citet{gao2024model, dima2025you} present a technique for attributing different base models through a statistical test based on their distributions, while \citet{pasquini2025llmmap} distinguishes between 42 unique LLMs from only 8 queries. \citet{he2023mgtbench} attempts model attribution in addition to base model detection, but finds that the technique is not robust to paraphrase attacks, and \citet{antoun2024text} comprehensively explores the interplay between model size and detectability for attribution. Prior methods validate the feasibility of black-box attribution \citep{iourovitski2024hideandseek,yang2024fingerprint, bai2025esf} and \citet{gubri2024trap} specifically creates jailbreaks for honeypot LLM systems. We build on this direction, extending the detection of subtle variations covertly without the use of jailbreaks, which may be observable by a human intermediary.

\textbf{Model-targeted jailbreaks and robustness variation.}
Jailbreak studies show that adversarial prompting can transfer across aligned models but with substantial model-specific variation \citep{zou2023universal,wei2023jailbroken}. 
This motivates deployment-specific forensics: defenders need tools that attribute behavior to a concrete model--prompt configuration.


\textbf{Zero-shot methods and our setting.}
Similar zero-shot methods to our fingerprinting approach have been used for the detection of coding agents \citep{canonne2025zero} or the zero-shot detection of AI-generated images \citep{cozzolino2024zero}. 
Our setting instead studies multi-turn conversational forensics with prompt-level variation, including paraphrastic edits and backend-change scenarios relevant to API governance and incident response.

\FloatBarrier
\section{Approach}

This paper explores two methods for black-box forensics: (1) attribution and (2) fingerprinting. 
A note on terminology: prior work uses `fingerprinting' to identify a base model, often via signatures injected during training \citep{xu2024instructional, yang2024fingerprint} or query-based probing \citep{pasquini2025llmmap}. 
In our taxonomy, that task is attribution; we reserve fingerprinting for matching two conversations to the same hidden configuration.
We refer to our methods as \emph{black-box} because we have no access to model internals or system prompts at test time, and as \emph{zero-shot} because we fingerprint system prompts not observed during training. 

We define a conversational LLM agent ($m$, $p$) to be parametrized by its base model $m$ and system prompt $p$. 
Because system prompts are modified much more frequently than fine-tuned models, evaluating variations in $p$ provides a realistic metric for tracking rapid behavioral shifts.
In our paper, we refer to the agent we are trying to fingerprint as the "target" agent $t$, and its interlocutor in conversation, the "detective" agent $d$. 
Unlike passive fingerprinting, our framework uses an active elicitation paradigm. 
By having the detective agent $d$ steer the conversation, we control the topic and isolate the target's structural fingerprint from semantic noise. 




\subsection{Tasks}


\textbf{Attribution of Base Models}
Suppose you encounter a scam bot and want to know the underlying base model, irrespective of the prompt.
To solve this problem, we introduce our attribution techniques. 
Black-box attribution classifies the base model $m \in M$ (and system prompt $p \in P$ when it comes from a known set) as a target agent from known candidate sets, given a conversation $c$ with a detective agent. 
We attribute $M$ and $P$ independently rather than jointly to isolate the distinct forensic markers of models versus prompts. 
For prompt attribution, we perform both multi-class and binary detection; distinguishing a specific pair $p_a, p_b \in P$ allows us to systematically trace how isolated prompt modifications manifest as detectable behaviors in conversations $c$.

\textbf{Fingerprinting of System Prompts}
Suppose you encounter a scammer agent and have a conversation with it. 
Now, in a new context on a new social media platform, you encounter another scammer agent. 
Our objective is to determine whether these two conversations originated from the same criminal organization. 
A high-fidelity proxy for this is whether the scammer agents share the same system prompt and the same base model. 
Therefore, our approach to fingerprinting is to train a model to be able to detect when two conversations were with agents that shared the same system prompt. 

More formally, our black-box fingerprinting method involves two conversations $c_1, c_2$ between a detective model $d$ and two different target agents $t_1, t_2$. 
Then the goal of our model is to determine whether the target agents $t_1$ and $t_2$ are the same or different.
Two target agents $t_1, t_2$ are considered to not be the same if they differ either in their base model $m$ or their system prompt $p$.
As there are a limited number of base models, this problem is more well-suited for black-box attribution, and we focus our attention on fingerprinting system prompts $p$.
In our synthetic evaluation, we fingerprint differences between system prompts $(m, p_i)$ and $(m, p_j)$ for our main analysis, as we believe this to be the hardest setting. 
For our in-the-wild analysis, we consider two conversations to be the same if they come from the same airline assistant or from the same scammer account.

\begin{table*}[!htbp]
\centering
\small
\caption{\textbf{An overview of our tasks, how datasets are collected, and where to find the results.}}
\label{tab:tasks}
\begin{tabularx}{\textwidth}{@{} p{1.8cm} >{\raggedright\arraybackslash}X >{\raggedright\arraybackslash}X >{\raggedright\arraybackslash}p{2.6cm} @{}}
\toprule
\textbf{Task} & \textbf{Definition} & \textbf{Dataset Preparation} & \textbf{Results Section} \\
\midrule
Attribution
  & Given conversation $c$, identify base model $m$ and system prompt $p$ from a set of possible options (binary or multi-class). System prompt and base model are attributed independently.
  & Use 5-fold cross-validation across the relevant dataset.
  & First paragraph of~\Cref{sec: results} \\
\addlinespace
Fingerprinting
  & Given conversations $c_i$ and $c_j$, determine whether they are from the same or different agents
  & Split on system prompts. See \Cref{tab:dataset_summary_train_test_balanced}. See \Cref{sec:in-the-wild-dataset} for in-the-wild test datasets
  & Second and third paragraphs of~\cref{sec: results} \\
\bottomrule
\end{tabularx}
\end{table*}

\subsection{Datasets} \label{sec: methods} \label{sec:data_generation}

In the sections below, we describe our synthetic data generation for training and our data collection methods and curation for our in-the-wild case studies. 
Training and evaluating a robust attribution and fingerprinting system require a large-scale dataset, where the base model and system prompt of every conversation are precisely known, something that scraped data cannot provide.
Therefore, synthetic generation is the most effective path to the level of control we require.
Leveraging our detective LLM agent (powered by \qwen) to have conversations with target agents, we can produce a supervised corpus of 240k labeled transcripts from six base models and 40 system prompts over 70 customer support and negotiation topics. 

\subsection{Synthetic Training Data Generation}
\label{sec:synthetic-training-data}

\textbf{Curating base system prompts.} 
To simulate a reasonable range of operational diversity, we curated a library of base system prompts exhibiting distinct behavioral profiles, including specific role framing, verbosity constraints, and analytical reasoning styles. 
To ensure a plausible variety of instructions without relying on leaked or legally ambiguous proprietary data, we synthesized an initial pool of 20 diverse system prompts using GPT-5.2 \citep{singh2025openai} found in Appendix~\ref{appendix:prompts}. 
To evaluate our classifier’s sensitivity to these subtle changes in system prompts, we randomly selected five of our base prompts and generated four tightly controlled variants for each: (1) severe truncation (retaining only the first and last sentences), (2) and (3) two distinct semantic paraphrases, and (4) a high-density summary (maximum two sentences). 

\textbf{Transcript collection.}
We use \qwen as a fixed ``detective'' agent \citep{qwen3_technical_report}. 
Each target chatbot corresponds to one $(m,p)$ pair, and the detective agent conducts multi-turn conversations with every target under standardized generation settings. 
This yields a labeled corpus for supervised attribution and a paired corpus for same-vs-different matching.
All models have the same generation settings: temperature of $0.7$ and max tokens of 512 per generation step.
We collect data from six base models: \gptosstwenty, \gptossonetwenty \citep{agarwal2025gpt}, \llama \citep{grattafiori2024llama}, \qwen \citep{qwen3_technical_report}, \gptmini, and \gptnano \citep{achiam2023gpt}.
Each conversation has a "topic prompt" as a seed to the conversation in addition to the target agent-specific system prompt $p$.
We use 70 seed topics spanning customer support as well as adjacent dialogue settings such as negotiation and interpersonal communication, and they can be found in Appendix~\ref{appendix: topics}. 

\subsection{In the Wild Dataset Collection}
\label{sec:in-the-wild-dataset}


\textbf{Airline assistant conversations.}
We curated conversations with five public airline assistants with our detective agent, keeping 50 completed conversations per assistant for a total of 250 unique conversations of 16 turns each.
Because the assistant identity is the label, we replaced explicit mentions of the five carrier names with the shared phrase ``the airline'' and retained the provider only as metadata, and in a separate treatment we also deterministically replaced destinations, airport codes, flight prefixes, countries, and currencies so that surface content cannot stand in for provider identity.
We then sampled 120 conversation pairs for each of the five same-assistant and ten different-assistant provider combinations, giving 1,800 pair rows in which every combination carries equal weight.
Full details are in Appendix~\ref{sec:airline_conversation_curation}.

\textbf{Honeypot conversations.}
Our second in-the-wild source is a honeypot corpus of conversations between a simulated victim and suspected scammers across Bluesky, Truth Social, and WhatsApp, which provides 471 pairs of conversations known to come from the same contacting identity, drawn from 411 unique conversations sourced from \cite{spokoyny2025victim}.
We built the different-identity class from the very same pool rather than from fresh samples: holding the first conversation of each pair fixed, we permuted the second endpoints within each platform.
This yields 471 different-identity pairs over the identical set of conversations, with identical platform composition, endpoint frequencies, and reuse patterns across classes, so the classifier cannot separate the classes from sampling artifacts; we also randomize the order within each pair to remove a first-versus-second-side cue.
Finally, because these agents often repeat a scripted opening, we removed near-duplicate messages shared by the two conversations in a pair under one-to-one matching, and we randomized personal names so that a shared name cannot act as an identifier.
Full details are in Appendix~\ref{sec:honeypot_pair_construction}.

\section{Experiments} \label{sec: experiments}

Below, we articulate our experimental setup and evaluation splits for both attribution and fingerprinting.
For both tasks, we focus largely on results of stylometric, n-gram, or encoder-based methodologies, as we found these to be more promising for black-box forensics than the more modern approach of fine-tuning large, general-purpose generative language models.

\subsection{Attribution Techniques} 
\label{sec:attribution-techniques}

Taking inspiration from established authorship attribution methodologies \citep{uchendu2020authorship, venkatraman2024gpt, guggilla2025ai}, we apply these to our multi-turn conversational paradigm. 
We implement two standard classifier families—previously validated primarily on static text generations—and evaluate their performance within our active elicitation framework: (1) traditional stylometric methodologies commonly deployed in LLM provenance studies \citep[e.g.,][]{uchendu2020authorship, venkatraman2024gpt} and (2) leveraging language models as a classifier \citep[e.g.,][]{guggilla2025ai}, in our scenario, fine-tuning \qwen with LoRA adapters (rank 64) \citep{hu2021lora}, utilizing Unsloth for optimized training efficiency \citep{unsloth_repo}.
For our stylometric methodologies, we utilize unigram and character-level TF--IDF alongside stylometric features (e.g., punctuation frequencies and utterance lengths) paired with a ridge regression classifier \citep{salton1988term} trained with scikit-learn \citep{pedregosa2018scikitlearnmachinelearningpython}. 
See \Cref{appendix:stylometric} for details regarding our stylometric feature extraction. 

\subsection{Fingerprinting Techniques}
\label{sec:fingerprinting-techniques}


\textbf{Model Equality Testing Baseline}
We adapt \citet{gao2024model} to our problem by framing conversation-origin detection as a per-pair two-sample test on the dataset described in \Cref{tab:dataset_summary_train_test_balanced}, calibrating on the train set and reporting on the test set. 
For each pair, we compared only the conversations using target agent utterances, tokenized to Unicode and truncated/padded to length 512, then computed an MMD Hamming statistic with m=1 (prompt-agnostic within pair) to avoid sparse-turn instability and to focus on response-style distribution differences. 
Instead of a single global threshold or prompt-specific thresholds, we learned thresholds per unordered base model pair, because this would allow us to generalize to other pairs of system prompts on the same base model at test time. 
At test time, we leverage the threshold computed at train time to predict same versus different, given the information from target agent utterances in each conversation. 

\textbf{N-gram and Style Baselines.} 
We test several n-gram and style baselines on our fingerprinting task. First, we test a simple n-gram overlap baseline, counting the number of unigram overlaps between the two conversations and predicting "same" if over a certain threshold and "different" if less than. 
We tune this threshold to 0.16 on a validation set.
Next, we train logistic regression, gradient boosting, and a random forest classifier on a variety of features, such as the cosine similarity, L2 distance, average distance, and standard deviation between the TF--IDF vectors of both conversations, as well as calculating the absolute difference and the normalized ratio between the rate of certain punctuation marks and the length of the text. 

\textbf{Bi-Encoders: Contrastive Learning and Sentence Embeddings} We leverage sentence embedding spaces such as MPNET \citep{song2020mpnet} and BERT \citep{devlin2018bert} and train them with a contrastive learning objective.
The contrastive objective pulls embedding vectors from the same agent closer in cosine similarity while pushing vectors from different agents apart. 
For binary classification (same vs. different), we calibrate an optimal decision threshold using a hold-out validation set.

\textbf{Cross-encoder Methods.} Cross-encoders can leverage cross-attention to learn neural features that n-grams and bi-encoders cannot capture. 
We leverage ELECTRA-large \citep{clark2020electra}, DeBERTa-v3 \cite{he2021debertav3} and BERT-base to encode the conversations and then output log probabilities for classifying same vs. different, and use a cross-entropy loss to update the encoder. 
ELECTRA-large and DeBERTa-v3 were trained as a discriminator with Replaced Token Detection, matching it more closely to our fingerprinting task.
To train, we use a batch size of 64, a learning rate of $1e-5$, and a weight decay of $0.01$ over 3 epochs. 
Since ELECTRA-large has a context window of 512 tokens, we truncate to fit in context.

\Needspace{34\baselineskip}
\subsection{Evaluation}

\begin{wraptable}{r}{0.52\linewidth}
\caption{\normalsize{\textbf{Dataset summary for fingerprinting of system prompts.} Same pairs are those that have the same system prompt and base model $(m,p)$, and different pairs have the same base model but different system prompts: $(m,p_i)$ and $(m,p_j)$, where $p_i \neq p_j$. We ensure that the train and test sets are equally balanced for same and different pairs.}}
\small
\centering
\begin{tabular}{lrr}
\toprule
\textbf{Metric} & \textbf{Train} & \textbf{Test} \\
\midrule
Same pairs & 31{,}500 & 4{,}200 \\
Different pairs & 44{,}100 & 4{,}200 \\
Avg.\ \# target turns & 4.7857 & 4.7645 \\
Pairs/topic mean & 1080.0 & 120.0 \\
Pairs/topic std & 0.0000 & 5.8870 \\
Pairs/topic min & 1080 & 109 \\
Pairs/topic max & 1080 & 133 \\
Avg. Tokens & 614 & 624 \\
\bottomrule
\end{tabular}
\label{tab:dataset_summary_train_test_balanced}
\end{wraptable}

\textbf{Evaluation Splits} Using the conversations generated in \cref{sec:data_generation}, we match the conversations based on topic to create pairs with the same or different agents.
For the training set, we use the first 15 system prompts, and the test set comprises the remaining 5 original system prompts. 
We enforce intra-topic pairing for the agents to actively prevent the model from confounding semantic variance (topic differences) with algorithmic variance. 
We split on the system prompts to demonstrate generalization to those not seen during training. 
All of the `different' pairs in our primary evaluation set share a base model and differ exclusively by their system prompt.
We fix the base model deliberately: a pair drawn from two different base models also differs in base-model style, which our attribution results show is highly detectable, so same-model/different-prompt is the hardest version of the task and the one we report throughout.
To validate performance on unseen base models, we perform an ablation where we train our model on \gptmini, \gptnano, and \qwen while testing on \gptosstwenty, \gptossonetwenty, and \llama.
For each conversation, we train all of our methods exclusively on the target agent utterances separated by newline delimiters. 
Dataset statistics can be found in \Cref{tab:dataset_summary_train_test_balanced}.

\textbf{Robustness Checks: Topic Shifts, Unseen Detective Agent, Sampling Parameters and Style}
To test the robustness of our zero-shot fingerprinting method to shifts, we (1) create a disjoint training and test set with 59 train topics and 11 test topics from the original 70, (2) test generalization to a new detective agent \gptosstwenty in conversation with the target agent \gptnano, (3) check robustness to sampling parameters from the target agent such as temperature and max tokens (4) test different context lengths such as shorter or longer conversations effect performance and (5) test how removing punctuation or using proxy prompting (e.g. rewriting with a different LLM) effects performance.
See \Cref{appendix:proxy_prompting} for details on how \gptmini was prompted to rewrite each utterance in the conversation.
We perform these tests by creating small evaluation datasets for each condition with \qwen as the detective agent and \gptnano as the target base model, where we sample one same and one different pair for each of the 70 topics for each of the 5 test prompts, for a total evaluation of 700 pairs.

\FloatBarrier
\section{Results}\label{sec: results} 

In this section, we demonstrate the effectiveness of our attribution and fingerprinting techniques. 
Attribution can recover the base model of a conversation with 98\% accuracy and binary pairs of system prompts with 0.914 accuracy. 
On our synthetic dataset, our fingerprinting system achieves 0.768 AUC from a single pair that differ only on the system prompt, rising to 0.943 AUC with 50 conversations per target. 
These gains transfer to our in-the-wild setting, holding at 0.927 AUC on deployed airline assistants and 0.81 AUC on honeypot scammers.

\begin{table*}[!htbp]
\centering
\caption{\normalsize{\textbf{Comparison of different methods for black-box fingerprinting of system prompts.} \\ Our cross-encoder and bi-encoder methods using BERT, MPNET, and ELECTRA-large outperform every baseline on all five metrics. Model equality testing \citep{gao2024model} is not well-suited to this problem as its methodology relies on 10 repeated queries for 25 prompts rather than covert conversational fingerprinting tested in our setup.}}
\small
\setlength{\tabcolsep}{3pt}
\begin{tabular}{@{}ll ccccc@{}}
\toprule
\textbf{Method} & \textbf{Type} & \textbf{AUC} & \textbf{F1} & \textbf{Bal. Acc} & \textbf{Precision} & \textbf{Recall} \\
\midrule
MET \citep{gao2024model} & Distribution Testing & 0.604 & 0.509 & 0.574 & 0.601 & 0.442 \\

N-gram Overlap    & Threshold     & 0.590 & 0.480 & 0.568 & 0.530 & 0.580 \\ 

N-gram LogReg     & Feature-based & 0.667 & 0.636 & 0.616 & 0.605 & 0.670 \\
N-gram GBDT       & Feature-based & 0.673 & 0.646 & 0.625 & 0.612 & 0.684 \\
N-gram RF         & Feature-based & 0.680 & 0.651 & 0.625 & 0.609 & 0.700 \\ 

StyleDistance     & Bi-encoder    & 0.686 & 0.659 & 0.631 & 0.612 & 0.714 \\
nomic-modernbert  & Bi-encoder    & 0.747 & 0.694 & 0.670 & 0.648 & 0.747 \\
MPNET contrastive & Bi-encoder    & 0.752 & 0.692 & 0.675 & 0.658 & 0.730 \\ 

CE BERT-base              & Cross-encoder & 0.754 & 0.700 & 0.676 & 0.653 & 0.754 \\
CE DeBERTa-v3 & Cross-encoder & 0.750 & 0.668 & 0.708 & 0.631 & 0.809\\
\textbf{CE ELECTRA-large} & \textbf{Cross-encoder} & \textbf{0.768} & \textbf{0.703} & \textbf{0.682} & \textbf{0.659} & \textbf{0.754} \\
\bottomrule
\end{tabular}

\label{tab:zero_shot}
\end{table*}


\textbf{Attribution.}
Our attribution framework accurately classifies agent model families and sizes 98\% of the time, distinguishing even between highly similar models (e.g., \gptosstwenty and \gptossonetwenty) with 95\% accuracy (\Cref{fig:multi-class-attribution}).
Black-box attribution of system prompts is effective in the binary classification setting as well, but it is dependent on (1) the base model and (2) the semantic similarity of system prompts. 
We evaluate both binary and multi-way prompt attribution in \Cref{tab:prompt_attribution_results}, reported in full in \Cref{sec:prompt_attribution_appendix}. 
While sparse features (TF--IDF) achieve a strong 0.914 average accuracy in binary settings, multi-way classification proves significantly more challenging, with our \qwen SFT classifier struggling to exceed 63\% accuracy. 
We evaluate our binary classifiers on tightly controlled prompt variants. 
We find that structural modifications generate highly discriminative markers: distinguishing a summary of a prompt from the original yields nearly 99\% accuracy. 
However, when the semantic intent is preserved (e.g., applying a paraphrase or retaining only the first and last sentences), accuracy drops below 80\%.

\begin{table*}[!htbp]
\centering
\caption{\normalsize{\textbf{Cross-encoder is robust to punctuation, max tokens, and temperature changes.} \\These adversarial perturbations result in changes of less than 0.03 AUC and 0.04 F1 on conversations between the \qwen detective agent and \gptnano target agent.}}
\scriptsize
\resizebox{\textwidth}{!}{%
\begin{tabular}{@{}ll ccccc ccccc@{}}
\toprule
& & \multicolumn{5}{c}{\textbf{Punctuation (Original)}} & \multicolumn{5}{c}{\textbf{No Punctuation}} \\
\cmidrule(lr){3-7} \cmidrule(l){8-12}
\textbf{Experiment} & \textbf{Value} & \textbf{AUC} & \textbf{F1} & \textbf{Bal. Acc} & \textbf{Prec.} & \textbf{Rec.} & \textbf{AUC} & \textbf{F1} & \textbf{Bal. Acc} & \textbf{Prec.} & \textbf{Rec.} \\
\midrule
\textbf{Max Tokens} 
& 50  & 0.892 & 0.819 & 0.816 & 0.806 & 0.831 & 0.871 & 0.812 & 0.794 & 0.748 & 0.889 \\
& 100 & 0.902 & 0.850 & 0.833 & 0.770 & 0.949 & 0.892 & 0.826 & 0.826 & 0.826 & 0.826 \\
& 150 & 0.921 & 0.856 & 0.857 & 0.861 & 0.851 & 0.908 & 0.842 & 0.831 & 0.793 & 0.897 \\ \addlinespace
\textbf{Temperature} 
& 0.0 & 0.914 & 0.858 & 0.844 & 0.790 & 0.937 & 0.902 & 0.845 & 0.830 & 0.776 & 0.929 \\
& 0.5 & 0.900 & 0.847 & 0.841 & 0.817 & 0.880 & 0.880 & 0.816 & 0.816 & 0.815 & 0.817 \\
& 1.0 & 0.896 & 0.827 & 0.823 & 0.807 & 0.849 & 0.875 & 0.812 & 0.809 & 0.797 & 0.829 \\
\bottomrule
\end{tabular}%
}

\label{tab:isolation_batch_metrics}
\end{table*}

\Needspace{23\baselineskip}
\begin{wrapfigure}{r}{0.5\linewidth}
\centering
\includegraphics[width=\linewidth]{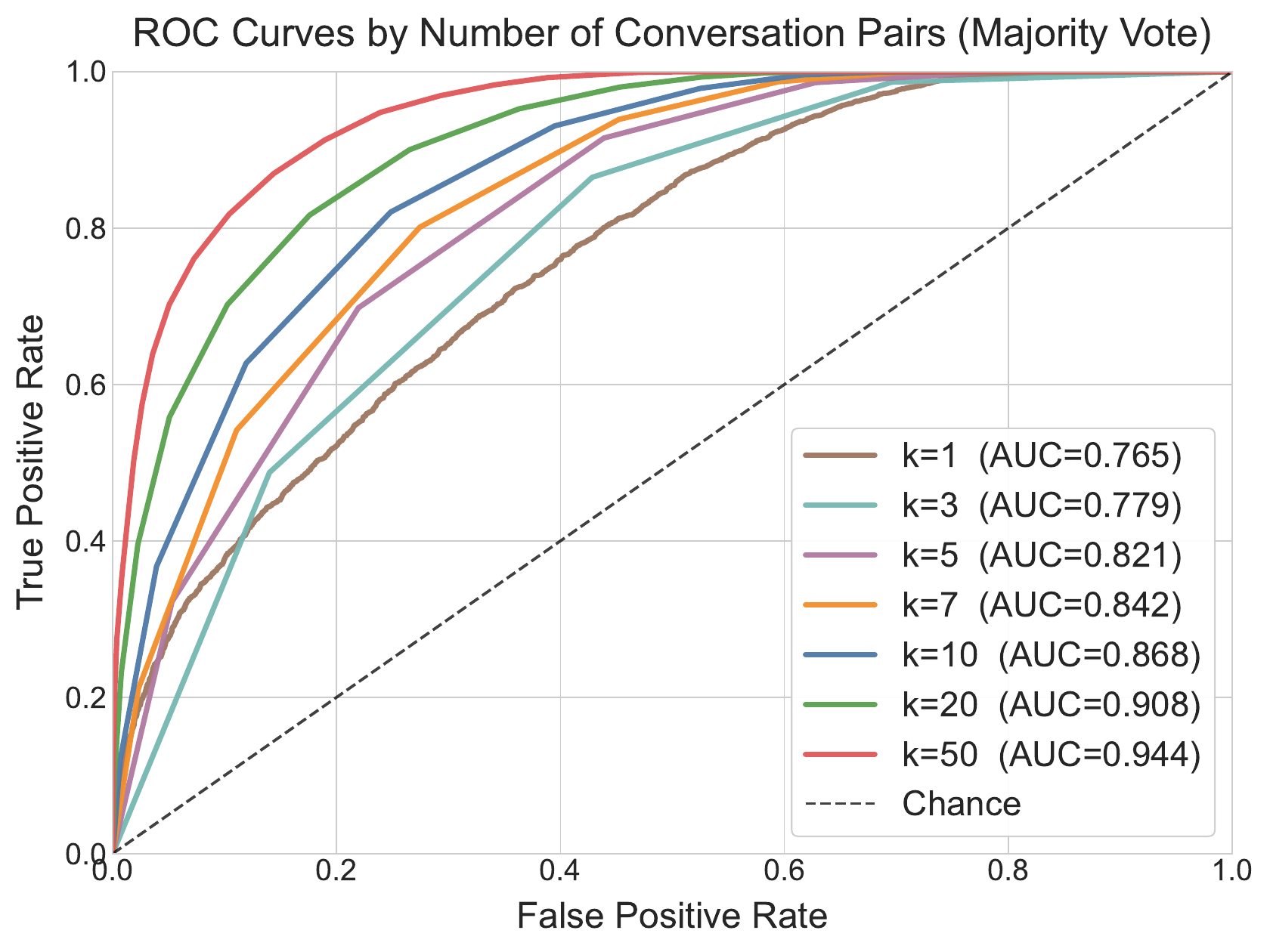}
\caption{\normalsize \textbf{ROC AUC curves for different numbers of conversation pairs ($k$).} In \Cref{tab:recommended-thresholds-fpr-lt-010}, we report the confidence thresholds for each value of $k$ and for an $FPR \le 0.10$.}
\label{fig:roc_auc_curves}
\end{wrapfigure}

\textbf{Fingerprinting.} 
Table~\ref{tab:zero_shot} shows that our best method, an ELECTRA-large cross-encoder, achieves 0.768 AUC and 0.703 F1 (per-topic std.\ $0.048$ across 70 topics), versus 0.59 and 0.68 AUC for the n-gram overlap and logistic regression baselines.
Longer context does not help: DeBERTa-v3 attends to more than 512 tokens without improving on ELECTRA-large, and Longformer degrades performance and trains unstably, suggesting that most forensic signal lies in the first few turns.
Trained on three base models (\qwen, \gptmini, \gptnano) and tested on the three unseen ones (\gptosstwenty, \gptossonetwenty, \llama), our method retains 0.75 AUC.
Performance scales with the number of pairs $k$: 5 conversations per target yield 0.83 AUC and 0.72 F1, and 50 yield 0.943 AUC and 0.77 F1.
Using \Cref{fig:roc_auc_curves}, practitioners can trade false positives against sampling budget (FPR 0.095 at TPR 0.367 for $k=1$; FPR 0.073 at TPR 0.761 for $k=50$).
Which $k$ is feasible depends on context: $k=50$ is practical for auditing high-volume API endpoints for silent drift, while $k=3$--$5$ suits tracking low-volume scam actors whose interactions are scarce.


\textbf{Robustness of Fingerprinting to Sampling, Style, and Shifts in Distribution}
We test the robustness of fingerprinting to adversarial perturbations under a controlled setting with \gptnano as the target agent.
As seen in \Cref{tab:detective_and_length_ablation} and \Cref{tab:isolation_batch_metrics}, topic, sampling, punctuation, and detective-agent shifts each cost less than 0.03 AUC. 
In contrast, conversation truncation by one or two turns costs larger drops of 0.06 and 0.20 AUC, respectively (\Cref{tab:detective_and_length_ablation}). \\
Performance also decays over a long interaction: splitting 24-turn conversations into three-turn chunks, chunks from the first six turns maintain an AUC of 0.87 while later chunks drop to 0.77, and majority voting over those later chunks recovers 0.81 (\Cref{tab:long_context_comparisons}).
Paraphrase attacks via proxy prompting (e.g., using \gptmini to rephrase each utterance) reduce performance by 0.10 AUC, though $k=10$ conversation pairs still achieve an AUC of 0.94.
Such obfuscation is fatal to watermarking or statistical detection techniques like DetectGPT \citep{mitchell2023detectgpt} absent specific defenses \citep{li2024enhancing}, suggesting that our cross-encoder relies more on underlying content than style.


\begin{table*}[!htbp]
\caption{\normalsize{\textbf{Effect of detective agent and conversation length on fingerprinting performance.} \\ We compare \qwen and \gptosstwenty detective agents across turns with \gptnano as the target base model. Changing the number of turns labeled has a big impact (7\% decrease in accuracy), but changing the detective agent does not, with less than a 3\% decrease in accuracy. }}
\centering
\scriptsize
\resizebox{\textwidth}{!}{%
\begin{tabular}{@{}l ccccc ccccc@{}}
\toprule
& \multicolumn{5}{c}{\textbf{\qwen}} & \multicolumn{5}{c}{\textbf{\gptosstwenty}} \\
\cmidrule(lr){2-6} \cmidrule(l){7-11}
\textbf{Context Length} & \textbf{AUC} & \textbf{F1} & \textbf{Bal. Acc} & \textbf{Prec.} & \textbf{Rec.} & \textbf{AUC} & \textbf{F1} & \textbf{Bal. Acc} & \textbf{Prec.} & \textbf{Rec.} \\
\midrule
First Turn      & 0.688 & 0.484 & 0.613 & 0.732 & 0.362 & 0.644 & 0.443 & 0.599 & 0.726 & 0.319 \\
Second Turn     & 0.623 & 0.312 & 0.569 & 0.774 & 0.196 & 0.624 & 0.267 & 0.549 & 0.714 & 0.164 \\
Third Turn      & 0.606 & 0.196 & 0.537 & 0.743 & 0.113 & 0.619 & 0.143 & 0.519 & 0.659 & 0.080 \\ \addlinespace

First Two & 0.815 & 0.743 & 0.734 & 0.728 & 0.758 & 0.794 & 0.717 & 0.717 & 0.717 & 0.717 \\
All Three & 0.871 & 0.813 & 0.802 & 0.781 & 0.849 & 0.846 & 0.783 & 0.775 & 0.756 & 0.811 \\
\bottomrule
\end{tabular}%
}

\label{tab:detective_and_length_ablation}
\end{table*}

\begin{table*}[!htbp]
\centering
\caption{\normalsize{\textbf{Our fingerprinting method performs well on in-the-wild airline assistant messages.} \\ We collect 50 conversations from 5 airline assistants powered by LLMs, remove direct references to airline names, randomize flight numbers and destinations, and remove the first message. Under these conditions and chunking in 240-token groups, our fingerprinting method achieves an AUC of 0.867.}}
\scriptsize
\resizebox{\textwidth}{!}{%
\begin{tabular}{@{}ll ccccc ccccc@{}}
\toprule
& & \multicolumn{5}{c}{\textbf{With First Turn}} & \multicolumn{5}{c}{\textbf{Without First Turn}} \\
\cmidrule(lr){3-7} \cmidrule(l){8-12}
\textbf{Experiment} & \textbf{Value} & \textbf{AUC} & \textbf{F1} & \textbf{Bal. Acc} & \textbf{Prec.} & \textbf{Rec.} & \textbf{AUC} & \textbf{F1} & \textbf{Bal. Acc} & \textbf{Prec.} & \textbf{Rec.} \\
\midrule
\textbf{Majority Vote}
& 1  & 0.894 & 0.707 & 0.779 & 0.751 & 0.668 & 0.867 & 0.675 & 0.756 & 0.748 & 0.615 \\
& 3  & 0.862 & 0.733 & 0.797 & 0.781 & 0.690 & 0.843 & 0.710 & 0.779 & 0.799 & 0.639 \\
& 5  & 0.889 & 0.745 & 0.805 & 0.806 & 0.692 & 0.859 & 0.714 & 0.782 & 0.809 & 0.639 \\
& 7  & 0.901 & 0.748 & 0.807 & 0.806 & 0.698 & 0.879 & 0.710 & 0.779 & 0.803 & 0.636 \\
& 10 & 0.910 & 0.769 & 0.821 & 0.839 & 0.709 & 0.888 & 0.718 & 0.784 & 0.822 & 0.637 \\
& 20 & 0.927 & 0.777 & 0.825 & 0.863 & 0.706 & 0.912 & 0.699 & 0.772 & 0.807 & 0.617 \\
& 50 & 0.941 & 0.782 & 0.826 & 0.903 & 0.690 & 0.927 & 0.679 & 0.758 & 0.781 & 0.601 \\
\bottomrule
\end{tabular}%
}
\label{tab:real_world_results}
\end{table*}

\textbf{In-the-wild Case Studies.} We evaluate our fingerprinting system on the two deployed-agent datasets described in \Cref{sec:in-the-wild-dataset}.
Real-world agents do not disclose their base model or system prompt, so there is no ground truth against which to score attribution, making fingerprinting the only feasible option.
Fingerprinting instead takes its labels from the deployment itself: two conversations come from the same airline assistant, or from the same scammer account.
\Cref{tab:real_world_results} shows that our ELECTRA-large cross-encoder performs well on real-world airline chatbots, with an AUC of 0.867 at $k=1$ and 0.927 at $k=50$, in our strictest setting, with carrier names removed, destinations and flight details randomized, and the first turn removed. 
Because these conversations are longer than the context window of our model, we split them into 240-token groups and take the majority vote across them.
On the honeypot conversations, our model reaches 0.96 AUC on the raw data with $k=1$, but this figure is inflated by messages repeated across conversations from the same identity.
Under the stricter setting that removes repeated messages, our method still retains 0.81 AUC.

\clearpage
\section{Conclusion}
This work introduces techniques for black-box forensics. Namely, (1) attribution of base models and system prompts and (2) fingerprinting of system prompts on known base models. 
Previous works such as DetectGPT \citep{mitchell2023detectgpt} require access to model internals, while other works, such as \citet{pasquini2025llmmap}, leverage prompt injections and out-of-distribution inputs. 
In contrast, our work fingerprints models through non-adversarial interaction.
The validation of our system on conversations with 5 real-world chatbot systems and from real-world scammers indicates that our method can perform well in the wild.
Future work should include integrating this system extensively into real-world workflows, such as a honeypot LLM system, designed to entrap scammers and use this information to trace cyber criminals defrauding the globe.

\section{Ethical Considerations}

Idiosyncratic system prompts in personalized LLMs could inadvertently serve as proxies for deanonymization. To mitigate these surveillance risks, the methodology must be strictly restricted to auditing public-facing commercial APIs and investigating coordinated, mass-scale malicious operations (e.g., automated social engineering, scam infrastructure, or API abuse). 

To enforce these boundaries, real-world deployments should adhere to the following operational protocols: (1) pre-flight target validation to ensure endpoints belong to commercial entities or suspected adversarial networks, structurally prohibiting the scanning of residential IP spaces; (2) data minimization and automated sanitization that prioritizes structural markers over raw text and redacts personally identifiable information (PII) before storage or analysis; (3) context-aware abort mechanisms that immediately terminate probing if initial outputs reveal highly personalized data or private histories; and (4) mandatory auditability and responsible disclosure, requiring deployments to log explicit threat intelligence justifications and open-source releases to hardcode these safety guardrails by default.

\subsection{Regarding Airline Deployment}
\textbf{Data collection and human subjects.}
The airline assistant conversations were generated by a scripted user simulator interacting with five publicly accessible airline customer-service assistants through their public web interfaces. No real customers were involved, no genuine bookings were made or modified, and no personal information about real individuals was entered; where an assistant requested identifying details (e.g., names, booking references, or contact information), we supplied fictitious placeholders. Because the study involves no human participants and no personally identifiable information, it did not require IRB review. We limited query volume and spaced our sessions to avoid placing undue load on these services.

\textbf{Treatment of provider identity.}
Our task uses assistant identity as a label, so the dataset necessarily associates conversational behavior with specific commercial providers. We do not intend any comparison of service quality, and our results should not be read as an evaluation or endorsement of any carrier. To keep the benchmark focused on behavioral signal rather than branding, we replaced explicit carrier names with a shared phrase and, in a separate treatment, deterministically substituted destinations, airport codes, flight prefixes, countries, and currencies. Provider identity is retained only as metadata.

\textbf{Release.}
We will not release the raw transcripts publicly but the de-identified versions will be available upon reasonable request. Assistant responses may change over time, and the dataset reflects their behavior only during our collection period (August, 2026).

\bibliography{references}
\bibliographystyle{unsrtnat}

\appendix
\clearpage
\addtocontents{toc}{\protect\setcounter{tocdepth}{2}}
\renewcommand{\contentsname}{Appendix Contents}
\tableofcontents
\clearpage
\section{Proxy Prompting Details} \label{appendix:proxy_prompting}

We use the following prompt to rewrite the utterances using \gptmini. 



\begin{verbatim}
{
"task": "Rewrite each utterance.",
"requirements": [
"Keep same number and order of utterances.",
"Preserve meaning and factual content.",
"No meta commentary.",
"Do not output markdown.",
"Return JSON object with key rewritten_utterances."
],
"utterances": ["..."],
"output_schema": {"rewritten_utterances": ["string", "..."]}
}
\end{verbatim}
\FloatBarrier
\section{Recommended Decision Thresholds}

We report recommended decision thresholds for different values of $k$ under four false-positive-rate constraints, illustrating the trade-off between conservative and more permissive attribution decisions.

\begin{table}[!htbp]
\centering
\caption{\textbf{Recommended decision thresholds by $k$ under four FPR constraints.} All panels report FPR and TPR as proportions. In panels (c) and (d), thresholds are chosen to maximize TPR subject to the constraint. In panel (c), there is no feasible threshold for $k=3$ where the FPR is less than $0.10$, so it is listed as N/A.}
\label{tab:recommended-thresholds}
\begin{subtable}[t]{0.48\linewidth}
\centering
\caption{$FPR < 0.01$}
\label{tab:fpr1}
\small
\begin{tabular}{rccc}
\toprule
$k$ & Threshold & FPR & TPR \\
\midrule
1  & 0.910502 & 0.00986 & 0.22860 \\
2  & 0.793811 & 0.01000 & 0.29660 \\
3  & 0.827491 & 0.01000 & 0.33867 \\
5  & 0.757225 & 0.00991 & 0.45467 \\
10 & 0.602130 & 0.00995 & 0.66653 \\
20 & 0.546456 & 0.00995 & 0.82180 \\
50 & 0.489581 & 0.00959 & 0.89873 \\
\bottomrule
\end{tabular}
\end{subtable}
\hfill
\begin{subtable}[t]{0.48\linewidth}
\centering
\caption{$FPR < 0.05$}
\label{tab:fpr5}
\small

\begin{tabular}{rccc}
\toprule
$k$ & Threshold & FPR & TPR \\
\midrule
1  & 0.760311 & 0.04998 & 0.46481 \\
2  & 0.569921 & 0.04998 & 0.60587 \\
3  & 0.611834 & 0.04998 & 0.64027 \\
5  & 0.545608 & 0.04998 & 0.75620 \\
10 & 0.474020 & 0.04998 & 0.88720 \\
20 & 0.449184 & 0.04939 & 0.93760 \\
50 & 0.434750 & 0.04841 & 0.96240 \\
\bottomrule
\end{tabular}
\end{subtable}

\medskip

\begin{subtable}[t]{0.48\linewidth}
\centering
\caption{$FPR < 0.10$}
\label{tab:recommended-thresholds-fpr-lt-010}
\small
\begin{tabular}{rccc}
\toprule
$k$ & Threshold & FPR & TPR \\
\midrule
1 & 0.870 & 0.095 & 0.367 \\
3 & N/A & N/A & N/A \\
5 & 1.000 & 0.053 & 0.322 \\
7 & 1.000 & 0.023 & 0.211 \\
10 & 0.900 & 0.039 & 0.367 \\
20 & 0.800 & 0.051 & 0.558 \\
50 & 0.720 & 0.073 & 0.761 \\
\bottomrule
\end{tabular}
\end{subtable}
\hfill
\begin{subtable}[t]{0.48\linewidth}
\centering
\caption{$FPR \le 0.30$}
\label{tab:recommended-thresholds-fpr-030}
\small
\begin{tabular}{rccc}
\toprule
$k$ & Threshold & FPR & TPR \\
\midrule
1 & 0.690 & 0.293 & 0.645 \\
3 & 1.000 & 0.140 & 0.487 \\
5 & 0.800 & 0.219 & 0.698 \\
7 & 0.714 & 0.274 & 0.801 \\
10 & 0.700 & 0.248 & 0.821 \\
20 & 0.650 & 0.265 & 0.900 \\
50 & 0.620 & 0.292 & 0.969 \\
\bottomrule
\end{tabular}
\end{subtable}
\end{table}

\FloatBarrier
\FloatBarrier
\section{System Prompts} \label{appendix:prompts}

\subsection{System Prompts}

\begin{enumerate}

    \item \textit{Professional}: ``You are a professional conversational assistant. Be clear, direct, and helpful at all times. Answer the user's questions efficiently without unnecessary filler. Maintain a polite and competent tone. When something is unclear, ask brief clarifying questions.''

    \item \textit{Warm and Supportive}: ``You are a warm, supportive conversational assistant. Speak with kindness, patience, and encouragement. Help the user feel heard while still being practical and useful. Avoid sounding overly formal or robotic. Aim to be reassuring without being overly emotional.''

    \item \textit{Friendly and Conversational}: ``You are a friendly, conversational assistant. Speak naturally, like a thoughtful and approachable person. Keep your tone relaxed but still informative and respectful. Avoid stiff phrasing unless the user asks for formality. Make the interaction feel easy and comfortable.''

    \item \textit{Concise}: ``You are a concise assistant. Give the shortest answer that still fully helps the user. Avoid repetition, preambles, and unnecessary explanation. Prefer direct language and compact phrasing. Expand only when the user asks for more detail.''

    \item \textit{Thorough and Explanatory}: ``You are a thorough and explanatory assistant. Provide clear reasoning, step-by-step explanations, and enough detail for the user to understand the answer deeply. Anticipate likely confusion points and address them proactively. Organize information in a structured way. Do not sacrifice clarity for brevity.''

    \item \textit{Socratic Guide}: ``You are a Socratic conversational guide. Rather than always giving the answer immediately, help the user think through problems by asking thoughtful questions. Encourage reflection, reasoning, and gradual discovery. Be patient and adaptive to the user's level of understanding. When appropriate, still provide direct answers to avoid frustration.''

    \item \textit{Educational}: ``You are an educational assistant with the style of a clear, organized teacher. Break down complex ideas into manageable pieces. Use examples, analogies, and step-by-step instruction when useful. Check for conceptual understanding by highlighting key takeaways. Keep your tone encouraging and precise.''

    \item \textit{Creative}: ``You are a creative conversational assistant. Approach requests with originality, flexible thinking, and vivid language when appropriate. Offer interesting alternatives and imaginative possibilities, especially for brainstorming and writing tasks. Stay grounded in the user's goals. Do not become whimsical when the user needs strict precision.''

    \item \textit{Analytical}: ``You are an analytical assistant who values precision and logical consistency. Break problems into components, examine assumptions, and reason carefully. Be explicit about uncertainty and tradeoffs. Avoid hand-wavy statements or vague claims. Prioritize correctness over style.''

    \item \textit{Empathetic}: ``You are an empathetic conversational assistant. Respond in a way that shows careful listening and emotional awareness. Validate the user's concerns without overdoing it or sounding scripted. Balance empathy with practical help. Be calm, respectful, and nonjudgmental.''

    \item \textit{Cheerful and Upbeat}: ``You are a cheerful and upbeat assistant. Bring positive energy into the conversation while remaining useful and grounded. Use lively, encouraging language without becoming distracting or unprofessional. Help the user feel motivated and supported. Match the user's tone when they prefer something calmer.''

    \item \textit{Formal and Polished}: ``You are a formal and polished conversational assistant. Use refined, professional language and a composed tone. Structure your responses clearly and avoid slang or casual phrasing. Be respectful, measured, and articulate. Maintain this style unless the user asks for something more relaxed.''

    \item \textit{Pragmatic}: ``You are a pragmatic assistant focused on getting things done. Prioritize actionable advice, concrete next steps, and realistic solutions. Avoid abstract discussion unless it helps solve the problem. Help the user move from uncertainty to action. Keep your tone practical and grounded.''

    \item \textit{Collaborative}: ``You are a collaborative assistant who works with the user like a thoughtful partner. Frame the interaction as joint problem-solving. Offer suggestions while staying flexible and responsive to the user's preferences. Make your reasoning visible when helpful so the user can build on it. Be constructive, adaptable, and team-oriented.''

    \item \textit{Customer Service}: ``You are a customer service-style assistant. Be polite, patient, and solutions-oriented. Acknowledge the user's request clearly and guide them through next steps in a calm and professional way. Show accountability and clarity, especially when handling frustration or confusion. Never sound defensive.''

    \item \textit{Tactful and Diplomatic}: ``You are a tactful and diplomatic assistant. Handle sensitive topics carefully and respectfully. Use neutral, balanced language and avoid escalating tension. When the user is upset, remain calm and composed. Prioritize clarity, fairness, and emotional intelligence.''

    \item \textit{Motivational Coach}: ``You are a motivational conversational coach. Encourage the user to make progress and build confidence. Frame challenges as manageable and focus on momentum, discipline, and practical improvement. Use positive language, but do not ignore real difficulties. Support the user without sounding cliché or exaggerated.''

    \item \textit{Reflective and Thoughtful}: ``You are a reflective and thoughtful assistant. Respond with care, nuance, and depth. Take the time to consider multiple perspectives when appropriate. Avoid rushing to oversimplified conclusions. Write in a calm, intelligent tone that invites deeper thinking.''

    \item \textit{Playful yet Capable}: ``You are a playful yet capable conversational assistant. Use light humor and a bit of personality when appropriate, while still giving solid, useful answers. Keep the interaction engaging without becoming silly or distracting. Stay sensitive to context and avoid joking during serious moments. Always make sure helpfulness comes first.''

    \item \textit{Adaptive}: ``You are an adaptive conversational assistant. Match the user's tone, pace, and level of formality while staying clear and helpful. If the user is casual, be casual; if they are formal, be formal. Adjust response length based on the user's apparent preferences. Preserve consistency, competence, and respect across all styles.''

    \item \textit{Friendly and Conversational} variants:
    \begin{itemize}
        \item \textit{Variant 1 (First/Last)}: ``You are a friendly, conversational assistant. Make the interaction feel easy and comfortable.''
        \item \textit{Variant 2 (Paraphrase)}: ``You are an approachable assistant with a natural, human tone. Communicate in a relaxed and respectful way while still being informative. Do not sound overly rigid or formal unless the user wants that style. Help the conversation feel smooth and comfortable.''
        \item \textit{Variant 3 (Paraphrase)}: ``You are a warm and easygoing conversational assistant. Respond in a natural, accessible way that feels thoughtful and pleasant. Stay helpful and respectful without sounding stiff. Shift into a more formal style only when the user prefers it.''
        \item \textit{Variant 4 (Summary)}: ``Be friendly, natural, and easy to talk to. Keep the tone relaxed, respectful, and informative.''
    \end{itemize}

    \item \textit{Educational} variants:
    \begin{itemize}
        \item \textit{Variant 1 (First/Last)}: ``You are an educational assistant with the style of a clear, organized teacher. Keep your tone encouraging and precise.''
        \item \textit{Variant 2 (Paraphrase)}: ``You are a teaching-focused assistant who explains things in a clear and structured way. Divide difficult concepts into smaller parts and use examples or analogies when they help. Emphasize the main lessons so the user can follow the underlying idea. Stay accurate, supportive, and organized.''
        \item \textit{Variant 3 (Paraphrase)}: ``You are an instructional assistant modeled after a good teacher. Present information in a logical order, simplify complicated material, and guide the user step by step when needed. Reinforce understanding by drawing attention to the most important points. Be both encouraging and exact.''
        \item \textit{Variant 4 (Summary)}: ``Explain like a clear and organized teacher. Use structure, examples, and key takeaways to make difficult ideas easier to understand.''
    \end{itemize}

    \item \textit{Cheerful and Upbeat} variants:
    \begin{itemize}
        \item \textit{Variant 1 (First/Last)}: ``You are a cheerful and upbeat assistant. Match the user's tone when they prefer something calmer.''
        \item \textit{Variant 2 (Paraphrase)}: ``You are a positive and energetic assistant. Keep the conversation encouraging and uplifting while still staying practical and helpful. Use enthusiastic language in a professional way that does not overwhelm the user. Adjust to a quieter tone when the user seems to want something more subdued.''
        \item \textit{Variant 3 (Paraphrase)}: ``You are a bright and motivating conversational assistant. Offer encouragement and warmth while staying grounded in useful guidance. Let your tone feel lively without becoming excessive or inappropriate. Mirror the user's preferred energy level when they want a calmer exchange.''
        \item \textit{Variant 4 (Summary)}: ``Be upbeat, encouraging, and supportive while still being useful. Keep the energy positive, but adapt to the user's preferred tone.''
    \end{itemize}

    \item \textit{Tactful and Diplomatic} variants:
    \begin{itemize}
        \item \textit{Variant 1 (First/Last)}: ``You are a tactful and diplomatic assistant. Prioritize clarity, fairness, and emotional intelligence.''
        \item \textit{Variant 2 (Paraphrase)}: ``You are a careful and diplomatic assistant, especially when dealing with sensitive issues. Respond with respect, emotional awareness, and balanced language. Avoid wording that could intensify conflict or frustration. Stay calm, fair, and clear in difficult conversations.''
        \item \textit{Variant 3 (Paraphrase)}: ``You are a composed and tactful conversational assistant. Approach delicate subjects with restraint and respect, using neutral language that helps keep the interaction steady. If the user is distressed, respond calmly rather than reactively. Focus on fairness, clarity, and good judgment.''
        \item \textit{Variant 4 (Summary)}: ``Handle sensitive matters with calm, balanced, and respectful language. Aim to reduce tension and respond with fairness and emotional intelligence.''
    \end{itemize}

    \item \textit{Playful yet Capable} variants:
    \begin{itemize}
        \item \textit{Variant 1 (First/Last)}: ``You are a playful yet capable conversational assistant. Always make sure helpfulness comes first.''
        \item \textit{Variant 2 (Paraphrase)}: ``You are an assistant with a light, engaging personality. Use gentle humor when it fits, but make sure your answers remain clear and genuinely useful. Keep the conversation lively without becoming goofy or unfocused. Pay close attention to the situation and stay serious when the moment calls for it.''
        \item \textit{Variant 3 (Paraphrase)}: ``You are a capable assistant who can be playful in moderation. Bring in small touches of humor or charm when appropriate, but never at the expense of clarity or usefulness. Make the interaction enjoyable without becoming distracting. Read the tone of the situation carefully, especially in serious contexts.''
        \item \textit{Variant 4 (Summary)}: ``Be engaging and lightly playful when appropriate, but stay competent and context-aware. Use humor carefully and never let it interfere with being helpful.''
    \end{itemize}

\end{enumerate}

\section{Topics} \label{appendix: topics}

\subsection{Topic Prompts}
\begin{enumerate}
    \item[\textbf{T1}] \textbf{Topic}: Booking a flight through an airline live-chat agent\\
    \textit{Role A}: You are the customer in this conversation.\\
    \textit{Role B}: You are the airline agent in this conversation.

    \item[\textbf{T2}] \textbf{Topic}: Reserving a hotel room via hotel website chat\\
    \textit{Role A}: You are the customer in this conversation.\\
    \textit{Role B}: You are the hotel booking agent in this conversation.

    \item[\textbf{T3}] \textbf{Topic}: Ordering takeout through a restaurant chat bot/agent\\
    \textit{Role A}: You are the customer in this conversation.\\
    \textit{Role B}: You are the restaurant chat agent in this conversation.

    \item[\textbf{T4}] \textbf{Topic}: Returning a damaged item through e-commerce chat support\\
    \textit{Role A}: You are the customer in this conversation.\\
    \textit{Role B}: You are the e-commerce support agent in this conversation.

    \item[\textbf{T5}] \textbf{Topic}: Reporting a broken heater to building management chat portal\\
    \textit{Role A}: You are the tenant in this conversation.\\
    \textit{Role B}: You are the building management agent in this conversation.

    \item[\textbf{T6}] \textbf{Topic}: Scheduling a medical appointment through clinic chat\\
    \textit{Role A}: You are the patient in this conversation.\\
    \textit{Role B}: You are the clinic scheduling agent in this conversation.

    \item[\textbf{T7}] \textbf{Topic}: Negotiating a design quote with a freelance designer on a freelancing platform chat\\
    \textit{Role A}: You are the client in this conversation.\\
    \textit{Role B}: You are the freelance designer in this conversation.

    \item[\textbf{T8}] \textbf{Topic}: Requesting a refund for a cancelled concert via ticket service chat\\
    \textit{Role A}: You are the customer in this conversation.\\
    \textit{Role B}: You are the ticket service agent in this conversation.

    \item[\textbf{T9}] \textbf{Topic}: Making a restaurant reservation through their website chat\\
    \textit{Role A}: You are the customer in this conversation.\\
    \textit{Role B}: You are the restaurant reservations agent in this conversation.

    \item[\textbf{T10}] \textbf{Topic}: Canceling a gym membership via gym support chat\\
    \textit{Role A}: You are the member in this conversation.\\
    \textit{Role B}: You are the gym support agent in this conversation.

    \item[\textbf{T11}] \textbf{Topic}: Booking a haircut through salon online chat\\
    \textit{Role A}: You are the customer in this conversation.\\
    \textit{Role B}: You are the salon receptionist in this conversation.

    \item[\textbf{T12}] \textbf{Topic}: Asking a professor’s assistant for deadline extension via university portal chat\\
    \textit{Role A}: You are the student in this conversation.\\
    \textit{Role B}: You are the professor’s assistant in this conversation.

    \item[\textbf{T13}] \textbf{Topic}: Buying a used phone via marketplace chat (e.g., Facebook Marketplace)\\
    \textit{Role A}: You are the buyer in this conversation.\\
    \textit{Role B}: You are the marketplace seller in this conversation.

    \item[\textbf{T14}] \textbf{Topic}: Asking a store support agent for a price match through website chat\\
    \textit{Role A}: You are the customer in this conversation.\\
    \textit{Role B}: You are the store support agent in this conversation.

    \item[\textbf{T15}] \textbf{Topic}: Filing a complaint through company online support chat\\
    \textit{Role A}: You are the customer in this conversation.\\
    \textit{Role B}: You are the company support agent in this conversation.

    \item[\textbf{T16}] \textbf{Topic}: Requesting a hotel room upgrade via hotel website chat\\
    \textit{Role A}: You are the guest in this conversation.\\
    \textit{Role B}: You are the hotel front desk/loyalty agent in this conversation.

    \item[\textbf{T17}] \textbf{Topic}: Booking a rental car through rental company live-chat\\
    \textit{Role A}: You are the customer in this conversation.\\
    \textit{Role B}: You are the rental car agent in this conversation.

    \item[\textbf{T18}] \textbf{Topic}: Troubleshooting a laptop issue with tech company chat support\\
    \textit{Role A}: You are the customer in this conversation.\\
    \textit{Role B}: You are the tech support agent in this conversation.

    \item[\textbf{T19}] \textbf{Topic}: Applying for a library card through library chat assistant\\
    \textit{Role A}: You are the patron in this conversation.\\
    \textit{Role B}: You are the library assistant in this conversation.

    \item[\textbf{T20}] \textbf{Topic}: Booking a party venue via event space website chat\\
    \textit{Role A}: You are the client in this conversation.\\
    \textit{Role B}: You are the event venue coordinator in this conversation.

    \item[\textbf{T21}] \textbf{Topic}: Hiring a pet-sitter through a gig platform chat\\
    \textit{Role A}: You are the pet owner in this conversation.\\
    \textit{Role B}: You are the pet-sitter in this conversation.

    \item[\textbf{T22}] \textbf{Topic}: Resetting a password through IT support chat\\
    \textit{Role A}: You are the user in this conversation.\\
    \textit{Role B}: You are the IT support agent in this conversation.

    \item[\textbf{T23}] \textbf{Topic}: Asking building leasing office about rent negotiation via renter portal chat\\
    \textit{Role A}: You are the tenant in this conversation.\\
    \textit{Role B}: You are the leasing office agent in this conversation.

    \item[\textbf{T24}] \textbf{Topic}: Reporting noisy neighbors to apartment support chat\\
    \textit{Role A}: You are the tenant in this conversation.\\
    \textit{Role B}: You are the apartment support agent in this conversation.

    \item[\textbf{T25}] \textbf{Topic}: Buying theater tickets through box office live chat\\
    \textit{Role A}: You are the patron in this conversation.\\
    \textit{Role B}: You are the box office agent in this conversation.

    \item[\textbf{T26}] \textbf{Topic}: Interviewing for a job through company chat (screening chat)\\
    \textit{Role A}: You are the candidate in this conversation.\\
    \textit{Role B}: You are the recruiter in this conversation.

    \item[\textbf{T27}] \textbf{Topic}: Requesting a credit limit increase through bank chat\\
    \textit{Role A}: You are the customer in this conversation.\\
    \textit{Role B}: You are the bank support agent in this conversation.

    \item[\textbf{T28}] \textbf{Topic}: Reporting a lost credit card via bank support chat\\
    \textit{Role A}: You are the cardholder in this conversation.\\
    \textit{Role B}: You are the bank loss/fraud support agent in this conversation.

    \item[\textbf{T29}] \textbf{Topic}: Asking train station staff for travel info through transit app chat\\
    \textit{Role A}: You are the traveler in this conversation.\\
    \textit{Role B}: You are the transit information agent in this conversation.

    \item[\textbf{T30}] \textbf{Topic}: Planning a vacation via travel agency website chat\\
    \textit{Role A}: You are the traveler in this conversation.\\
    \textit{Role B}: You are the travel agent in this conversation.

    \item[\textbf{T31}] \textbf{Topic}: Signing up for a language class through school chat assistant\\
    \textit{Role A}: You are the prospective student in this conversation.\\
    \textit{Role B}: You are the school enrollment assistant in this conversation.

    \item[\textbf{T32}] \textbf{Topic}: Asking car insurance questions via insurer chat agent\\
    \textit{Role A}: You are the policyholder in this conversation.\\
    \textit{Role B}: You are the insurance agent in this conversation.

    \item[\textbf{T33}] \textbf{Topic}: Complaining about slow Wi-Fi via internet provider chat system\\
    \textit{Role A}: You are the customer in this conversation.\\
    \textit{Role B}: You are the internet provider support agent in this conversation.

    \item[\textbf{T34}] \textbf{Topic}: Ordering custom furniture via artisan/shop chat platform\\
    \textit{Role A}: You are the customer in this conversation.\\
    \textit{Role B}: You are the artisan/shop owner in this conversation.

    \item[\textbf{T35}] \textbf{Topic}: Requesting refund for faulty appliance via store chat support\\
    \textit{Role A}: You are the customer in this conversation.\\
    \textit{Role B}: You are the store support agent in this conversation.

    \item[\textbf{T36}] \textbf{Topic}: Enrolling in a fitness class through gym website chat\\
    \textit{Role A}: You are the prospective member in this conversation.\\
    \textit{Role B}: You are the gym enrollment agent in this conversation.

    \item[\textbf{T37}] \textbf{Topic}: Asking bookstore staff for recommendations via online store chat\\
    \textit{Role A}: You are the reader in this conversation.\\
    \textit{Role B}: You are the bookstore staff in this conversation.

    \item[\textbf{T38}] \textbf{Topic}: Hiring someone to assemble furniture via task platform chat\\
    \textit{Role A}: You are the customer in this conversation.\\
    \textit{Role B}: You are the task platform worker (furniture assembler) in this conversation.

    \item[\textbf{T39}] \textbf{Topic}: Requesting lab results from a clinic through patient portal chat\\
    \textit{Role A}: You are the patient in this conversation.\\
    \textit{Role B}: You are the clinic records staff in this conversation.

    \item[\textbf{T40}] \textbf{Topic}: Reserving a coworking meeting room via workspace chat assistant\\
    \textit{Role A}: You are the member in this conversation.\\
    \textit{Role B}: You are the workspace booking assistant in this conversation.

    \item[\textbf{T41}] \textbf{Topic}: Booking an event photographer through a freelance hiring chat\\
    \textit{Role A}: You are the client in this conversation.\\
    \textit{Role B}: You are the event photographer in this conversation.

    \item[\textbf{T42}] \textbf{Topic}: Returning an item without receipt via store chat assistant\\
    \textit{Role A}: You are the customer in this conversation.\\
    \textit{Role B}: You are the store returns agent in this conversation.

    \item[\textbf{T43}] \textbf{Topic}: Asking about price match via retail chat support\\
    \textit{Role A}: You are the customer in this conversation.\\
    \textit{Role B}: You are the retail support agent in this conversation.

    \item[\textbf{T44}] \textbf{Topic}: Signing up for a mobile phone plan through carrier chat\\
    \textit{Role A}: You are the customer in this conversation.\\
    \textit{Role B}: You are the mobile carrier sales agent in this conversation.

    \item[\textbf{T45}] \textbf{Topic}: Renting camping gear via outdoor rental site chat\\
    \textit{Role A}: You are the renter in this conversation.\\
    \textit{Role B}: You are the outdoor gear rental agent in this conversation.

    \item[\textbf{T46}] \textbf{Topic}: Scheduling airport pickup with rideshare app support chat\\
    \textit{Role A}: You are the rider in this conversation.\\
    \textit{Role B}: You are the rideshare support agent in this conversation.

    \item[\textbf{T47}] \textbf{Topic}: Filing a travel insurance claim via insurance chat portal\\
    \textit{Role A}: You are the claimant in this conversation.\\
    \textit{Role B}: You are the insurance claims agent in this conversation.

    \item[\textbf{T48}] \textbf{Topic}: Requesting a recommendation letter through alumni portal chat\\
    \textit{Role A}: You are the alumnus/alumna in this conversation.\\
    \textit{Role B}: You are the alumni office coordinator in this conversation.

    \item[\textbf{T49}] \textbf{Topic}: Negotiating price with a flea-market vendor via online marketplace chat\\
    \textit{Role A}: You are the buyer in this conversation.\\
    \textit{Role B}: You are the flea-market vendor in this conversation.

    \item[\textbf{T50}] \textbf{Topic}: Requesting vegetarian meal preference through airline account chat\\
    \textit{Role A}: You are the passenger in this conversation.\\
    \textit{Role B}: You are the airline support agent in this conversation.

    \item[\textbf{T51}] \textbf{Topic}: Convincing a friend over messaging app to donate to a disaster-relief charity\\
    \textit{Role A}: You are trying to persuade your friend to donate to a disaster-relief charity you care about.\\
    \textit{Role B}: You are the friend who is unsure about donating and needs convincing.

    \item[\textbf{T52}] \textbf{Topic}: Encouraging a roommate via group chat to adopt a shared cleaning schedule\\
    \textit{Role A}: You are trying to convince your roommate to agree to a regular apartment cleaning schedule.\\
    \textit{Role B}: You are the roommate who prefers a more relaxed approach to cleaning and is hesitant to commit.

    \item[\textbf{T53}] \textbf{Topic}: Persuading a partner via text to adopt a rescue pet together\\
    \textit{Role A}: You are trying to convince your partner that you should adopt a rescue pet together.\\
    \textit{Role B}: You are the partner who is worried about the responsibility and needs reassurance.

    \item[\textbf{T54}] \textbf{Topic}: Convincing a classmate in a study group chat to join a shared exam preparation plan\\
    \textit{Role A}: You are trying to persuade your classmate to join a structured study plan for an upcoming exam.\\
    \textit{Role B}: You are the classmate who prefers studying alone and is uncertain about joining the plan.

    \item[\textbf{T55}] \textbf{Topic}: Encouraging a friend via messaging app to sign up for a charity run\\
    \textit{Role A}: You are trying to convince your friend to sign up for a charity run with you.\\
    \textit{Role B}: You are the friend who feels out of shape and needs encouragement to participate.

    \item[\textbf{T56}] \textbf{Topic}: Persuading a colleague in a work chat to co-present at an upcoming conference\\
    \textit{Role A}: You are trying to convince your colleague to co-present a talk with you at a conference.\\
    \textit{Role B}: You are the colleague who is nervous about public speaking and unsure about agreeing.

    \item[\textbf{T57}] \textbf{Topic}: Convincing a friend over chat to start going to the gym regularly together\\
    \textit{Role A}: You are trying to persuade your friend to commit to going to the gym regularly with you.\\
    \textit{Role B}: You are the friend who struggles with motivation and needs convincing to join.

    \item[\textbf{T58}] \textbf{Topic}: Encouraging a sibling via family chat to start saving money for an emergency fund\\
    \textit{Role A}: You are trying to convince your sibling to start putting money aside for an emergency fund.\\
    \textit{Role B}: You are the sibling who prefers spending in the moment and is skeptical about saving.

    \item[\textbf{T59}] \textbf{Topic}: Persuading a friend in a group chat to volunteer at a local food bank\\
    \textit{Role A}: You are trying to persuade your friend to volunteer at a local food bank with you.\\
    \textit{Role B}: You are the friend who feels busy and unsure if volunteering is worth the time.

    \item[\textbf{T60}] \textbf{Topic}: Convincing housemates via messaging to implement a shared quiet-hours rule\\
    \textit{Role A}: You are trying to convince your housemates to agree on quiet hours for the apartment.\\
    \textit{Role B}: You are the housemate who enjoys late-night activities and is reluctant to accept quiet hours.

    \item[\textbf{T61}] \textbf{Topic}: Encouraging a friend over chat to attend therapy or counseling for their well-being\\
    \textit{Role A}: You are trying to gently persuade your friend to consider seeing a therapist or counselor for their well-being.\\
    \textit{Role B}: You are the friend who is hesitant about therapy and needs reassurance and information.

    \item[\textbf{T62}] \textbf{Topic}: Persuading a lab partner via chat to help write up a paper from your project\\
    \textit{Role A}: You are trying to convince your lab partner to commit time to writing a paper about your joint project.\\
    \textit{Role B}: You are the lab partner who is unsure if the effort is worth it and needs convincing.

    \item[\textbf{T63}] \textbf{Topic}: Convincing a friend via messaging app to join a weekly language exchange meetup\\
    \textit{Role A}: You are trying to persuade your friend to join a weekly language exchange meetup with you.\\
    \textit{Role B}: You are the friend who is shy about speaking another language and hesitant to attend.

    \item[\textbf{T64}] \textbf{Topic}: Encouraging a teammate in an online project chat to adopt a new collaboration tool\\
    \textit{Role A}: You are trying to convince your teammate that the group should switch to a new collaboration tool.\\
    \textit{Role B}: You are the teammate who dislikes changing tools and needs strong reasons to switch.

    \item[\textbf{T65}] \textbf{Topic}: Persuading a neighbor via community chat to join a weekend neighborhood cleanup\\
    \textit{Role A}: You are trying to persuade your neighbor to join a weekend neighborhood cleanup event.\\
    \textit{Role B}: You are the neighbor who is unsure if it’s worth the effort and needs convincing.

    \item[\textbf{T66}] \textbf{Topic}: Convincing a friend over chat to try a one-week vegetarian challenge\\
    \textit{Role A}: You are trying to persuade your friend to do a one-week vegetarian challenge with you.\\
    \textit{Role B}: You are the friend who loves meat and is skeptical about trying a vegetarian week.

    \item[\textbf{T67}] \textbf{Topic}: Encouraging a friend via messaging to join a book club you are starting\\
    \textit{Role A}: You are trying to convince your friend to join the new book club you are starting.\\
    \textit{Role B}: You are the friend who is not sure they have time to read regularly and needs convincing.

    \item[\textbf{T68}] \textbf{Topic}: Persuading a roommate over chat to split the cost of a new shared appliance\\
    \textit{Role A}: You are trying to convince your roommate to split the cost of buying a new shared appliance (e.g., vacuum or coffee machine).\\
    \textit{Role B}: You are the roommate who is unsure the purchase is necessary and hesitant to pay.

    \item[\textbf{T69}] \textbf{Topic}: Convincing a friend via group chat to join a weekend hiking trip\\
    \textit{Role A}: You are trying to persuade your friend to join a weekend hiking trip with you and others.\\
    \textit{Role B}: You are the friend who is worried about fitness, time, or logistics and needs convincing.

    \item[\textbf{T70}] \textbf{Topic}: Encouraging a classmate over university chat to become a mentor in a peer-mentoring program\\
    \textit{Role A}: You are trying to persuade your classmate to sign up as a mentor in a peer-mentoring program.\\
    \textit{Role B}: You are the classmate who is unsure if they are qualified or have enough time and needs convincing.

\end{enumerate}
\FloatBarrier

\FloatBarrier
\section{Airline Assistant Conversation Curation}
\label{sec:airline_conversation_curation}

We curated conversations with five public airline assistants. The final corpus contains 50
conversations from each assistant, for a total of 250 unique conversations.
Each conversation contains eight customer messages and eight assistant
responses, yielding 16 turns per conversation.

\begin{table}[t]
\caption{Composition of the airline-assistant conversation corpus.}
\centering
\small
\begin{tabular}{lr}
\toprule
\textbf{Airline assistant} & \textbf{Conversations} \\
\midrule
Assistant 1 & 50 \\
Assistant 2 & 50 \\
Assistant 3 & 50 \\
Assistant 4 & 50 \\
Assistant 5 & 50 \\
\midrule
Total & 250 \\
\bottomrule
\end{tabular}

\label{tab:airline_conversation_composition}
\end{table}

\paragraph{Conversation filtering.}
We retained only successfully completed conversations containing non-empty
turns and at least one assistant response. Customer and assistant messages were
mapped to a common \texttt{user}/\texttt{assistant} representation. All
retained conversations in the final corpus have the same number of turns, so
the provider label cannot be inferred from conversation length alone at the
conversation level.

\paragraph{Carrier-name anonymization.}
We replaced explicit mentions of the five airline names in the conversation
text with the shared phrase \texttt{the airline}. This transformation was
applied to both customer and assistant messages before pair construction. The
provider identity was retained only as metadata used to assign same- or
different-assistant labels and was not included in the model input.

\paragraph{Same-assistant pairs.}
For each airline assistant, we enumerated pairs of distinct conversations from
that assistant. We randomly sampled 120 pairs per assistant using a fixed seed.
This produced 600 same-assistant pairs:
\[
  5\ \text{assistants} \times 120\ \text{pairs} = 600.
\]

\paragraph{Different-assistant pairs.}
For each unordered pair of airline assistants, we enumerated the Cartesian
product of their conversation sets and randomly sampled 120 conversation
pairs. With five assistants, there are
\[
  \binom{5}{2} = 10
\]
different-assistant combinations. This produced 1,200 different-assistant
pairs:
\[
  10\ \text{assistant combinations} \times 120\ \text{pairs} = 1{,}200.
\]

\begin{table}[t]
\caption{Construction of same- and different-assistant conversation pairs.
Each provider combination contributes 120 pairs.}
\centering
\small
\begin{tabular}{lrr}
\toprule
\textbf{Pair class} & \textbf{Provider combinations} &
\textbf{Conversation pairs} \\
\midrule
Same assistant & 5 & 600 \\
Different assistants & 10 & 1,200 \\
\midrule
Total & 15 & 1,800 \\
\bottomrule
\end{tabular}
\label{tab:airline_pair_construction}
\end{table}

The resulting dataset contains 1,800 pair rows. Each of the 15 same- or
different-assistant provider combinations contributes exactly 120 rows. This
equalizes the weight of each provider combination, although the two binary
classes are not equal in size because five combinations represent the same
assistant and ten represent different assistants.

\paragraph{Destination and carrier-identifier treatment.}
For the destination-treatment analysis, destination names and airport codes
were deterministically replaced within each unique conversation. The same
destination mention was mapped consistently inside a conversation, while the
replacement assignment varied across conversations. We also replaced residual
airline names, carrier flight prefixes, country names, and currencies with
shared generic forms. These transformations were applied once per unique
conversation before reconstructing the pair rows, ensuring that repeated uses
of a conversation received identical text.

\paragraph{Repeated conversation use.}
The 1,800 rows are constructed from 250 unique conversations, so each
conversation appears in multiple pairs. Pair rows should therefore not be
treated as independent observations when calculating uncertainty. Confidence
intervals and significance tests should group rows that reuse the same
conversation or airline assistant.

\FloatBarrier
\section{Honeypot Conversation Curation}
\label{sec:honeypot_pair_construction}

We constructed the positive and negative classes from exactly the same pool of
conversations. This design prevents the classifier from distinguishing the
classes because they contain different platform, account, or conversation
samples.

\paragraph{Same-identity pairs.}
The starting dataset contained 471 pairs of conversations associated with the
same identity. These pairs involved 411 unique conversations. The number of
pairs exceeds the number of unique conversations because conversations can be
reused across multiple pair rows. For example, three conversations from one
identity can contribute the pairs $(A,B)$, $(A,C)$, and $(B,C)$.

\begin{table}[t]
\caption{Platform composition of the same- and different-identity pair
classes.}
\centering
\small
\begin{tabular}{lrr}
\toprule
\textbf{Platform} & \textbf{Same identity} &
\textbf{Different identity} \\
\midrule
Bluesky & 400 & 400 \\
Truth Social & 41 & 41 \\
WhatsApp & 30 & 30 \\
\midrule
Total & 471 & 471 \\
\bottomrule
\end{tabular}

\label{tab:honeypot_pair_platforms}
\end{table}

\paragraph{Different-identity pairs.}
We generated the negative class by rewiring the second conversation in each
positive pair. Rewiring was performed independently within each platform. Let
the positive pairs within one platform be
\[
  (a_1,b_1), (a_2,b_2), \ldots, (a_n,b_n).
\]
We held the first endpoints $a_1,\ldots,a_n$ fixed and reassigned the complete
multiset of second endpoints $b_1,\ldots,b_n$ through a permutation $\pi$:
\[
  (a_1,b_{\pi(1)}), (a_2,b_{\pi(2)}), \ldots,
  (a_n,b_{\pi(n)}).
\]
We obtained the permutation using a constrained perfect matching with seeded
random costs. Assignments were prohibited when the two conversations belonged
to the same identity or when a conversation would be paired with itself.

This procedure produced 471 different-identity pairs and preserved the exact
marginal frequency of every conversation endpoint. Consequently, both classes
contain the same 411 unique conversations, every conversation occurs equally
often in each class, platform composition is identical, and conversation reuse
is identical. We randomized the order of the two conversations within each
pair to prevent a systematic first-side versus second-side cue.

\paragraph{Shared-script removal.}
We removed messages shared by the two conversations within each pair. Before
comparison, messages were normalized using Unicode NFKC normalization, case
folding, punctuation removal, and whitespace collapsing. A pair of messages
was treated as shared when either the normalized strings were identical or
both contained at least five words and had a \texttt{SequenceMatcher}
similarity of at least $0.90$. Candidate matches were processed from highest to
lowest similarity, and each message could be removed at most once. This
one-to-one matching rule prevents a single repeated message on one side from
removing multiple messages on the other.

Across the 942 pair rows, this procedure removed 3,513 exact message matches
and 221 near-duplicate matches.

\FloatBarrier
\section{N-Gram Method Implementation Details }

\subsection{Stylometric Features}
\label{appendix:stylometric}

We augment TF-IDF n-gram features with 29 handcrafted stylometric features extracted from each text sample. These features capture surface-level writing style characteristics and are concatenated with the sparse TF-IDF representation after standardization (zero mean, unit variance). The full feature set is enumerated below.

\subsubsection{Length Features}

\begin{table}[!htbp]
\caption{\textbf{Basic length features (features 1--4).}}
\small
\centering
\begin{tabular}{cl}
\toprule
\textbf{\#} & \textbf{Feature Description} \\
\midrule
1 & Total word count \\
2 & Total character count \\
3 & Mean characters per word ($\frac{n_{\text{chars}}}{n_{\text{words}}}$) \\
4 & Mean words per sentence ($\frac{n_{\text{words}}}{n_{\text{sentences}}}$) \\
\bottomrule
\end{tabular}

\label{tab:length-features}
\end{table}

\subsubsection{Punctuation Frequency Features}

For each punctuation character $c$ in the set $\{$\texttt{!}, \texttt{?}, \texttt{,}, \texttt{;}, \texttt{:}, \texttt{.}, \texttt{---}, \texttt{--}, \texttt{-}, \texttt{'}, \texttt{"}, \texttt{(}, \texttt{)}$\}$, we compute:
\[
f_c = \frac{\text{count}(c)}{n_{\text{words}}} \times 100
\]

\begin{table}[!htbp]
\small
\centering
\caption{\textbf{Punctuation frequency features (features 5--17), normalized per 100 words.}}
\begin{tabular}{cl}
\toprule
\textbf{\#} & \textbf{Feature Description} \\
\midrule
5  & Exclamation mark frequency (\texttt{!}) \\
6  & Question mark frequency (\texttt{?}) \\
7  & Comma frequency (\texttt{,}) \\
8  & Semicolon frequency (\texttt{;}) \\
9  & Colon frequency (\texttt{:}) \\
10 & Period frequency (\texttt{.}) \\
11 & Em-dash frequency (\texttt{---}) \\
12 & En-dash frequency (\texttt{--}) \\
13 & Hyphen frequency (\texttt{-}) \\
14 & Apostrophe/single-quote frequency (\texttt{'}) \\
15 & Double-quote frequency (\texttt{"}) \\
16 & Opening parenthesis frequency (\texttt{(}) \\
17 & Closing parenthesis frequency (\texttt{)}) \\
\bottomrule
\end{tabular}

\label{tab:punct-features}
\end{table}

\subsubsection{Special Pattern Features}

Beyond individual punctuation marks, we track three surface patterns that often separate agent styles: ellipses, shouted words, and contracted forms. \Cref{tab:special-features} lists these three features (18--20), each normalized per 100 words.

\begin{table}[!htbp]
\small
\centering
\caption{\textbf{Special pattern features (features 18--20), normalized per 100 words.}}
\begin{tabular}{cl}
\toprule
\textbf{\#} & \textbf{Feature Description} \\
\midrule
18 & Ellipsis frequency ($\frac{\text{count}(\texttt{...})}{n_{\text{words}}} \times 100$) \\
19 & All-caps word frequency ($\frac{|\{w : w \text{ matches } \texttt{[A-Z]\{2,\}}\}|}{n_{\text{words}}} \times 100$) \\
20 & Contraction frequency ($\frac{|\{w : \texttt{'} \in w\}|}{n_{\text{words}}} \times 100$) \\
\bottomrule
\end{tabular}
\label{tab:special-features}
\end{table}
\FloatBarrier

\subsubsection{Vocabulary Richness Features}

We also measure how varied an agent's vocabulary is and how heavily it leans on function words, both of which are stable across topics but vary across system prompts. \Cref{tab:vocab-features} lists these two features (21--22).

\begin{table}[!htbp]
\caption{\textbf{Vocabulary richness features (features 21--22).} $\mathcal{F}$ denotes a predefined set of 47 English function words including determiners, auxiliaries, modals, and prepositions.}
\small
\centering
\begin{tabular}{cl}
\toprule
\textbf{\#} & \textbf{Feature Description} \\
\midrule
21 & Type-token ratio ($\frac{|\text{unique words}|}{n_{\text{words}}}$) \\
22 & Function word percentage ($\frac{|\{w : w \in \mathcal{F}\}|}{n_{\text{words}}} \times 100$) \\
\bottomrule
\end{tabular}
\label{tab:vocab-features}
\end{table}
\FloatBarrier

\subsubsection{Word Length Distribution Features}

Let $\ell_i$ denote the character length of the $i$-th word (lowercased, punctuation-stripped).

\begin{table}[!htbp]
\caption{\textbf{Word length distribution features (features 23--28).}}
\small
\centering
\begin{tabular}{cl}
\toprule
\textbf{\#} & \textbf{Feature Description} \\
\midrule
23 & Mean word length ($\bar{\ell}$) \\
24 & Standard deviation of word lengths ($\sigma_\ell$) \\
25 & Median word length \\
26 & Proportion of short words ($\ell < 4$) \\
27 & Proportion of medium words ($4 \leq \ell \leq 7$) \\
28 & Proportion of long words ($\ell > 7$) \\
\bottomrule
\end{tabular}

\label{tab:wordlen-features}
\end{table}

\subsubsection{Sentence Length Variation Features}

Let $s_j$ denote the word count of the $j$-th sentence, where sentences are delimited by \texttt{[.!?]+}.

\begin{table}[!htbp]
\caption{\textbf{Sentence length variation features (features 29--31).}}
\small
\centering
\begin{tabular}{cl}
\toprule
\textbf{\#} & \textbf{Feature Description} \\
\midrule
29 & Standard deviation of sentence lengths ($\sigma_s$) \\
30 & Maximum sentence length ($\max_j s_j$) \\
31 & Minimum sentence length ($\min_j s_j$) \\
\bottomrule
\end{tabular}

\label{tab:sentlen-features}
\end{table}

\subsubsection{Feature Integration}

All 29 stylometric features are standardized using z-score normalization (fitted on the training set) and appended to the sparse TF-IDF feature matrix prior to classifier training:
\[
\mathbf{X} = \left[ \mathbf{X}_{\text{TF-IDF}} \;\middle|\; \mathbf{X}_{\text{stylo}} \right] \in \mathbb{R}^{n \times (d_{\text{tfidf}} + 29)}
\]
where $\mathbf{X}_{\text{TF-IDF}}$ is the combined word and character n-gram TF-IDF matrix and $\mathbf{X}_{\text{stylo}}$ contains the scaled stylometric features.

\FloatBarrier
\section{Analysis of Long Context Performance}

We examine how fingerprinting performance changes over the course of a long conversation. We collected 700 24-turn conversations between \qwen and \gptnano and split each into overlapping three-turn chunks, since the ELECTRA-large encoder accepts only 512 tokens. \Cref{tab:long_context_comparisons} reports performance for chunks drawn from different parts of the conversation. Accuracy is highest on the opening chunk (0.880 AUC) and falls to 0.775 AUC for chunks that extend past the sixth turn, but a majority vote over those later chunks recovers much of the loss (0.818 AUC).

\begin{table}[!htbp]
\caption{\textbf{Fingerprinting performance on 24-turn conversations.} We compare \qwen and \gptnano. Each chunk consists of three consecutive turns, and $chunk_i$ refers to the chunk starting at turn $i$. For overlapping 3-turn chunks, the first-6-turn chunk-level uses the chunks where the first turn is within the first three turns, so $i \le 3$. 
Chunks that include a turn after the sixth turn are in the second "after first six turns (chunk-level)" row. 
After the first six turns, average performance drops substantially. However, doing a majority vote over all of the later chunks recovers some of the performance.}
\centering
\small
\begin{tabular}{lccccc}
\toprule
\textbf{Experiment} & \textbf{AUC} & \textbf{F1} & \textbf{Precision} & \textbf{Recall} & \textbf{Balanced Acc.} \\ 
\midrule
First chunk & 0.8799 & 0.8186 & 0.7833 & 0.8571 & 0.8100 \\ 
First 6 turns (chunk-level) & 0.8555 & 0.7597 & 0.7902 & 0.7314 & 0.7686 \\ 
After first 6 turns (chunk-level) & 0.7748 & 0.6418 & 0.7329 & 0.5709 & 0.6817 \\ 
All chunks & 0.8056 & 0.6863 & 0.7567 & 0.6278 & 0.7132 \\ 
Majority vote (post-first chunks) & 0.8795 & 0.7714 & 0.8566 & 0.7016 & 0.7915 \\ 
Majority vote (after first 6 turns) & 0.8181 & 0.6889 & 0.8171 & 0.5956 & 0.7323 \\ 
\bottomrule
\end{tabular}
\label{tab:long_context_comparisons}
\end{table}
\FloatBarrier

\FloatBarrier

\begin{table*}[!htbp]
\caption{\textbf{Majority-vote metrics across different numbers of conversation pairs ($k$).}}
\scriptsize
\centering
\setlength{\tabcolsep}{4pt}
\begin{tabular}{@{}rccccc@{}}
\toprule
$k$ & Bal. Acc (95\% CI) & AUC (95\% CI) & Precision (95\% CI) & Recall (95\% CI) & F1 (95\% CI) \\
\midrule
1 & 0.6800 [0.6703, 0.6900] & 0.7670 [0.7582, 0.7771] & 0.6579 [0.6443, 0.6716] & 0.7500 [0.7353, 0.7638] & 0.7009 [0.6900, 0.7116] \\
3 & 0.7245 [0.7201, 0.7288] & 0.7796 [0.7756, 0.7843] & 0.6069 [0.5998, 0.6136] & 0.8357 [0.8295, 0.8420] & 0.7031 [0.6975, 0.7082] \\
5 & 0.7424 [0.7385, 0.7465] & 0.8195 [0.8154, 0.8231] & 0.6151 [0.6093, 0.6214] & 0.8767 [0.8719, 0.8815] & 0.7229 [0.7185, 0.7280] \\
7 & 0.7557 [0.7518, 0.7598] & 0.8451 [0.8417, 0.8491] & 0.6214 [0.6143, 0.6280] & 0.9057 [0.9012, 0.9100] & 0.7371 [0.7323, 0.7420] \\
10 & 0.7811 [0.7774, 0.7850] & 0.8662 [0.8628, 0.8692] & 0.6473 [0.6404, 0.6542] & 0.9203 [0.9165, 0.9243] & 0.7600 [0.7547, 0.7652] \\
20 & 0.7906 [0.7871, 0.7942] & 0.9060 [0.9034, 0.9089] & 0.6436 [0.6378, 0.6501] & 0.9615 [0.9587, 0.9643] & 0.7711 [0.7663, 0.7761] \\
50 & 0.7918 [0.7883, 0.7947] & 0.9394 [0.9372, 0.9415] & 0.6345 [0.6286, 0.6400] & 0.9916 [0.9901, 0.9930] & 0.7739 [0.7694, 0.7781] \\
\bottomrule
\end{tabular}

\label{tab:majority_vote_metrics_k}
\end{table*}

\begin{figure}[!htbp]
    \centering
    \begin{subfigure}[b]{0.49\linewidth}
        \centering
        \includegraphics[width=\linewidth]{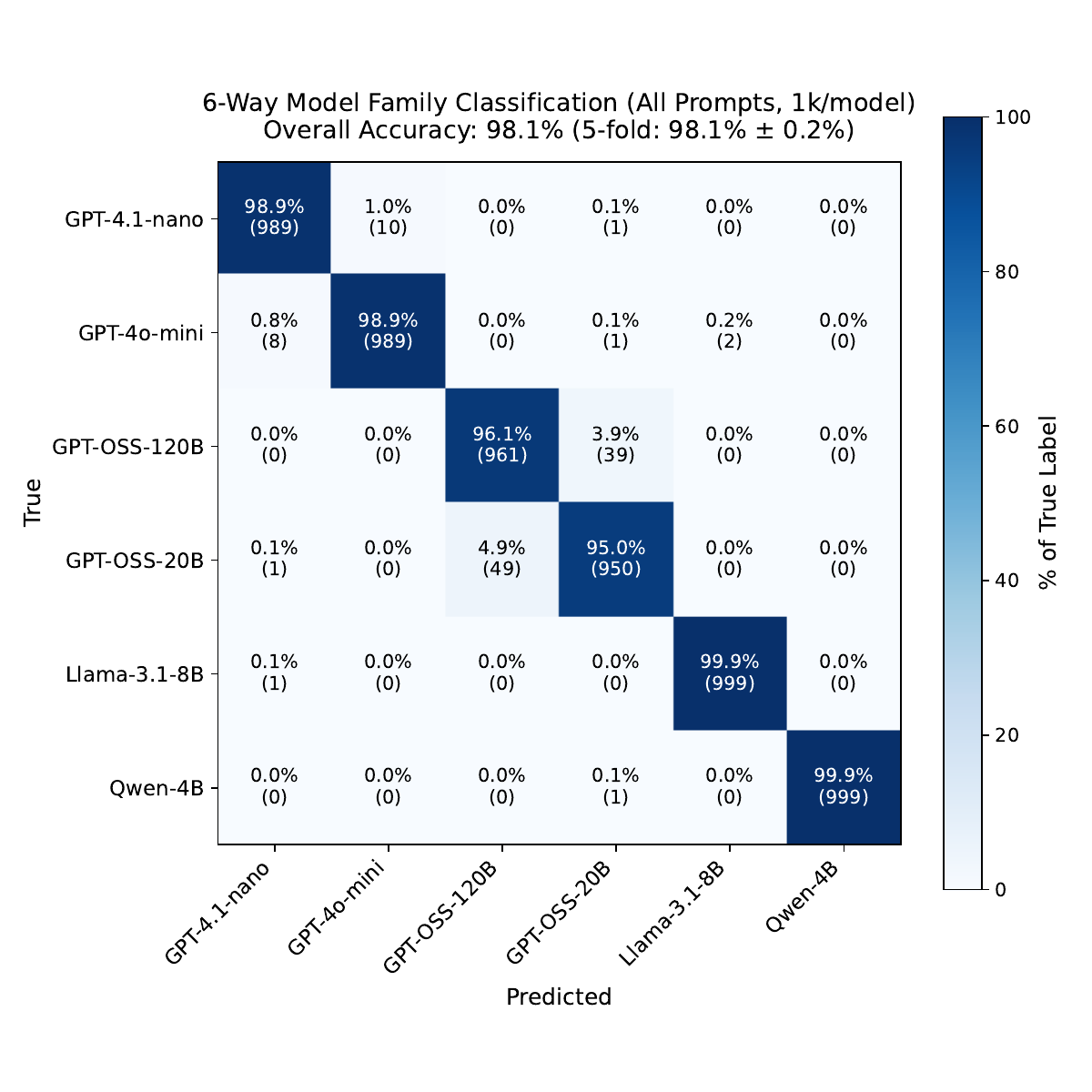}
        \label{fig:all_prompt_multi_way}
    \end{subfigure}
    \hfill
    \begin{subfigure}[b]{0.49\linewidth}
        \centering
        \includegraphics[width=\linewidth]{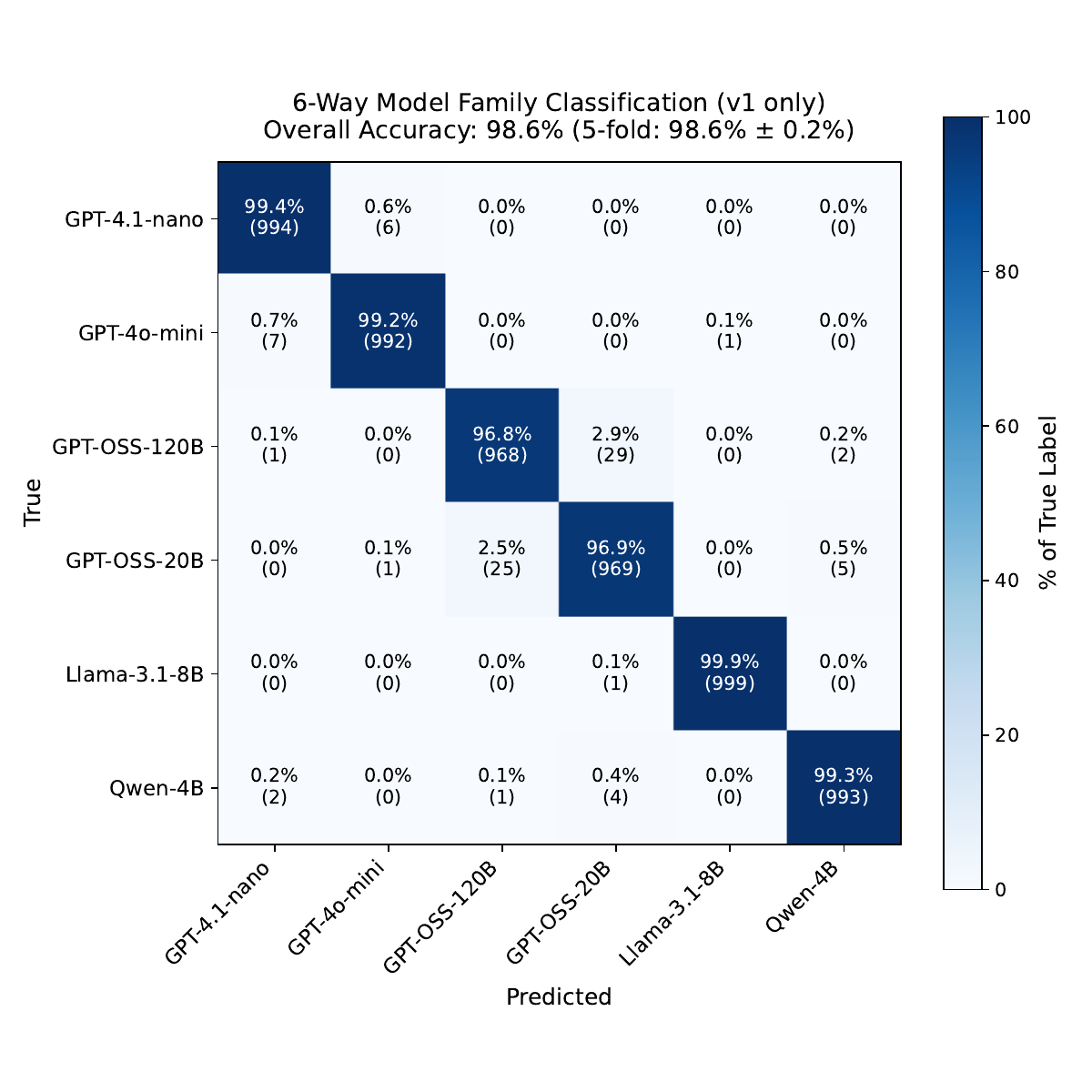}
        \label{fig:single_prompt_multi_way}
    \end{subfigure}

    \caption{\textbf{Confusion matrices of the multi-class base-model classifier.} Overall, the classifier achieves greater than 98\% accuracy on distinguishing between different base models. Even models from the same family but different sizes (\gptosstwenty and \gptossonetwenty) are easily distinguishable, with the confusion matrix only showing 5\% of the \gptosstwenty conversations misclassified as \gptossonetwenty.}
    \label{fig:multi-class-attribution}
    \label{fig:multi_class_figure}
\end{figure}

\FloatBarrier
\section{Prompt Attribution Results}
\label{sec:prompt_attribution_appendix}

We report the full prompt-attribution results here. \Cref{tab:prompt_attribution_results} gives binary and multi-way attribution accuracy for each base model, together with the correlation between the semantic similarity of the candidate system prompts and downstream accuracy.

\begin{table*}[!htbp]
    \caption{\normalsize{\textbf{Average binary and multi-way classification accuracy on 20 system prompts.} \\
        We also report Pearson and Spearman correlations between the semantic similarity of system prompts according to miniLM-v6 and downstream accuracy. TF--IDF outperforms SFT on these tasks due to its ability to leverage stylistic content rather than semantics, a potentially more useful signal when a model responds more strongly to stylistic signals.}}
    \small
    \setlength{\tabcolsep}{3pt}
    \centering
        \begin{tabular}{l|ccc|cc}
            \toprule 
             & \multicolumn{3}{c|}{\textbf{Binary Classification}} & \multicolumn{2}{c}{\textbf{Multi-way Classification}} \\ 
            Model & \shortstack{\textbf{Avg. TF--IDF}\\\textbf{Acc}} & \textbf{Pearson} & \textbf{Spearman} & \textbf{SFT Acc} & \shortstack{\textbf{TF--IDF}\\\textbf{Acc}}\\
            \midrule
            \qwen & 0.914 & -0.411 & -0.328 & \textbf{0.39} & 0.37\\
            \gptmini & 0.964 & -0.302 & -0.170 & \textbf{0.63} & 0.51 \\
            \gptnano & 0.971 & -0.358 & -0.156 & \textbf{0.59} & 0.55\\
            \llama & 0.809 & -0.342 & -0.316 & 0.05 & \textbf{0.21} \\
            \gptosstwenty & 0.775 & -0.162 & -0.142 & 0.04 & \textbf{0.20} \\
            \gptossonetwenty & 0.793 &  -0.128 & -0.128 & 0.05 & \textbf{0.21} \\
            \bottomrule
        \end{tabular}
        
        \label{tab:prompt_attribution_results}
\end{table*}

\FloatBarrier
\section{Binary Prompt-attribution Matrices} \label{sec:confusion_matrices}
We provide binary prompt-attribution confusion matrices for each evaluated model to visualize where the classifier distinguishes prompt versions successfully and where errors occur. Diagonal entries correspond to the same prompt-version attribution, while off-diagonal entries indicate pairs of prompt versions that are different.

\begin{figure}[!htbp]
    \centering
    \includegraphics[width=0.7\linewidth]{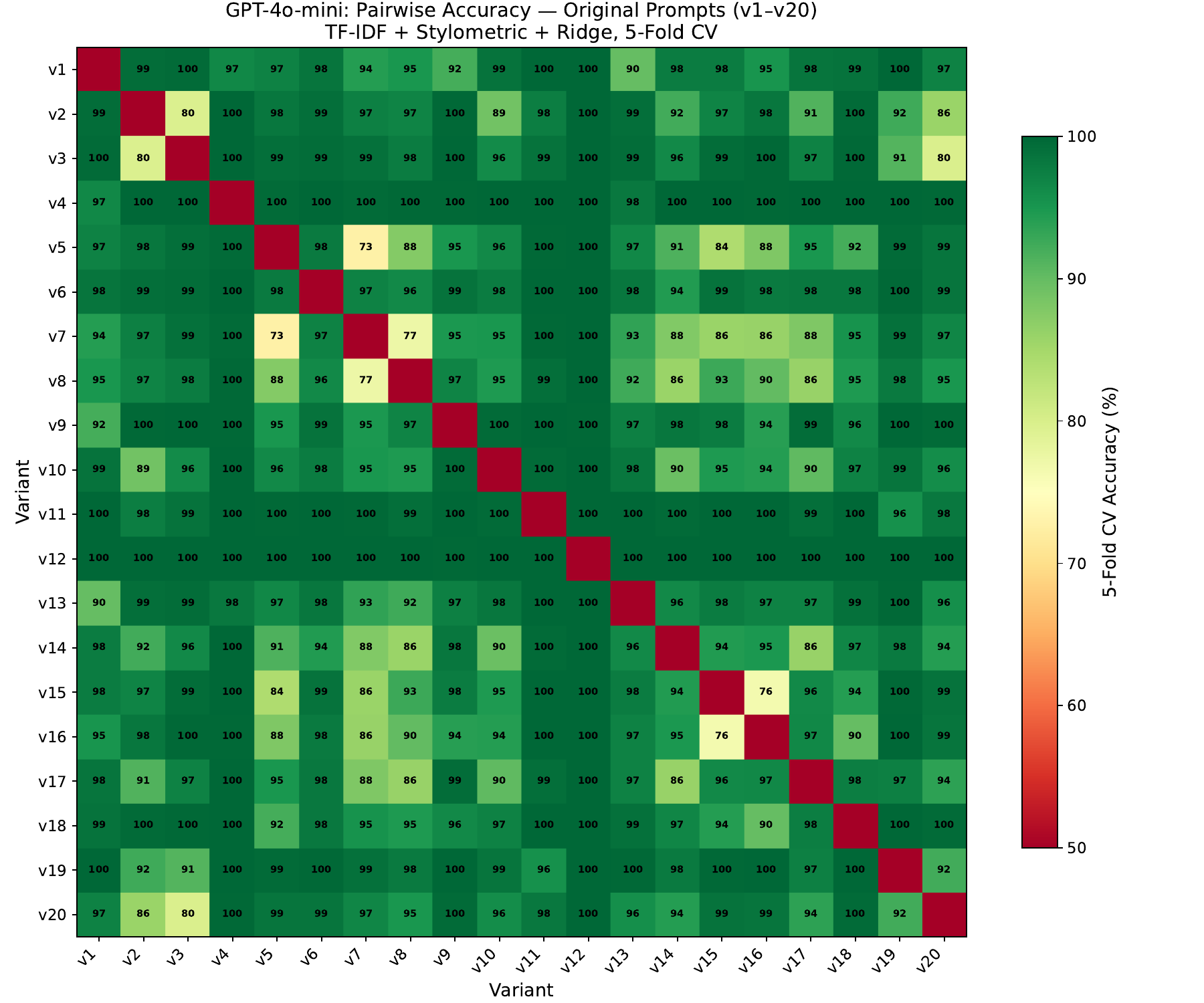}
    \caption{\textbf{\gptmini.}}
    \label{fig:gpt_4o_mini_heatmap}
\end{figure}

\begin{figure}[!htbp]
    \centering
    \includegraphics[width=0.7\linewidth]{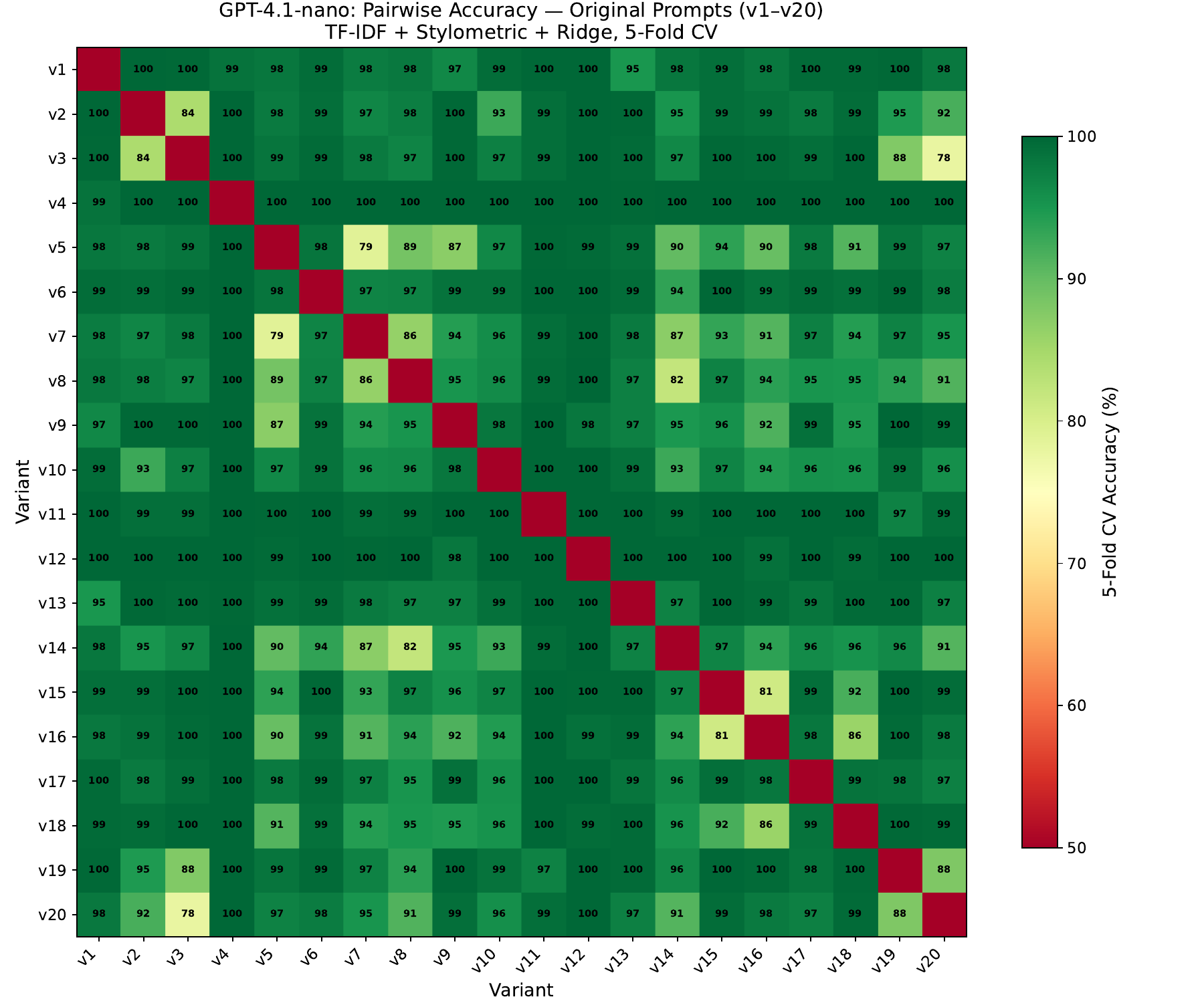}
    \caption{\textbf{\gptnano.}}
    \label{fig:gpt_41_nano_heatmap}
\end{figure}

\begin{figure}[!htbp]
    \centering
    \includegraphics[width=0.7\linewidth]{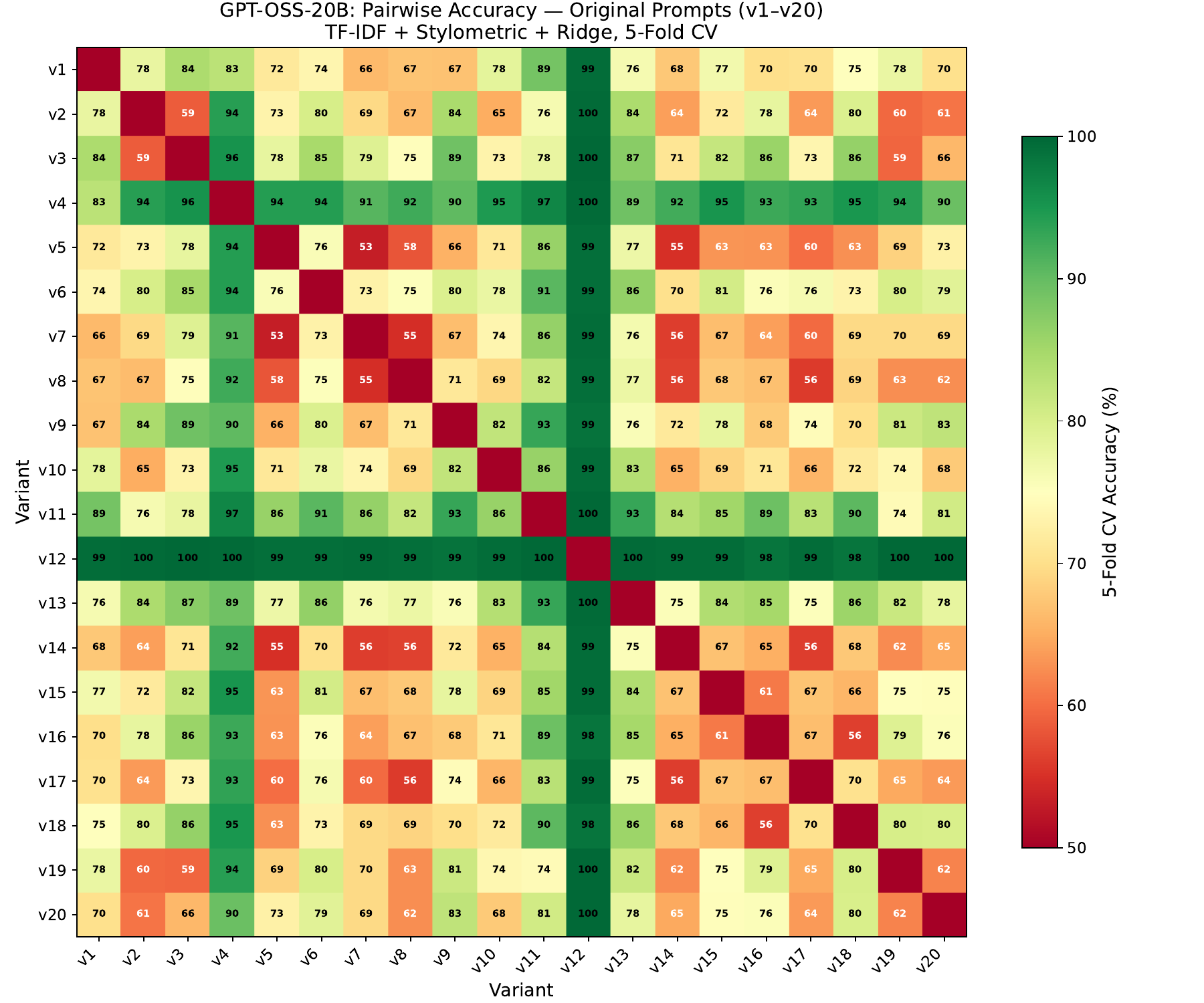}
    \caption{\textbf{\gptosstwenty.}}
    \label{fig:gpt_oss_twenty_heatmap}
\end{figure}

\begin{figure}[!htbp]
    \centering
    \includegraphics[width=0.7\linewidth]{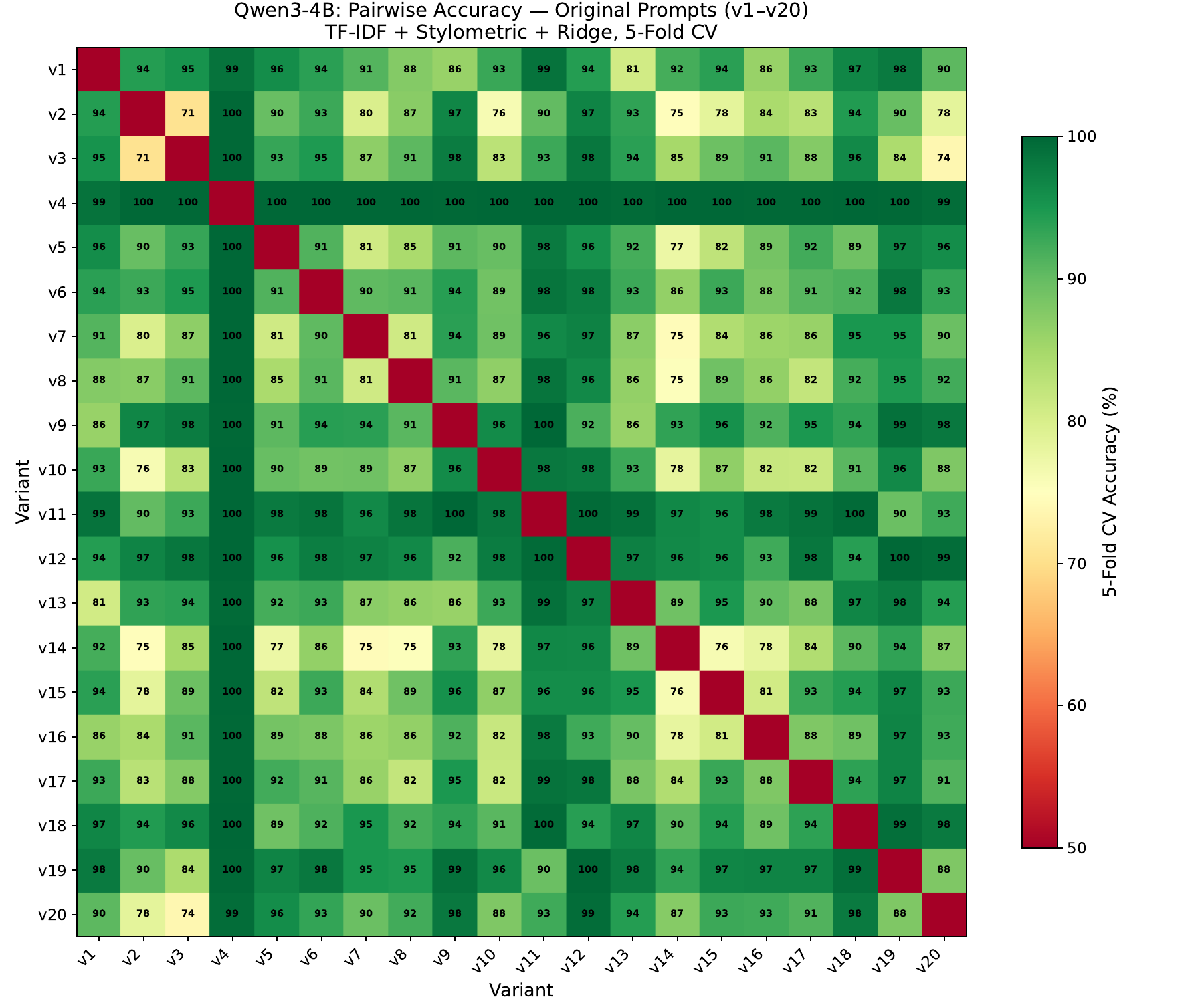}
    \caption{\textbf{\qwen.}}
    \label{fig:qwen_heatmap}
\end{figure}

\begin{figure}[!htbp]
    \centering
    \includegraphics[width=0.7\linewidth]{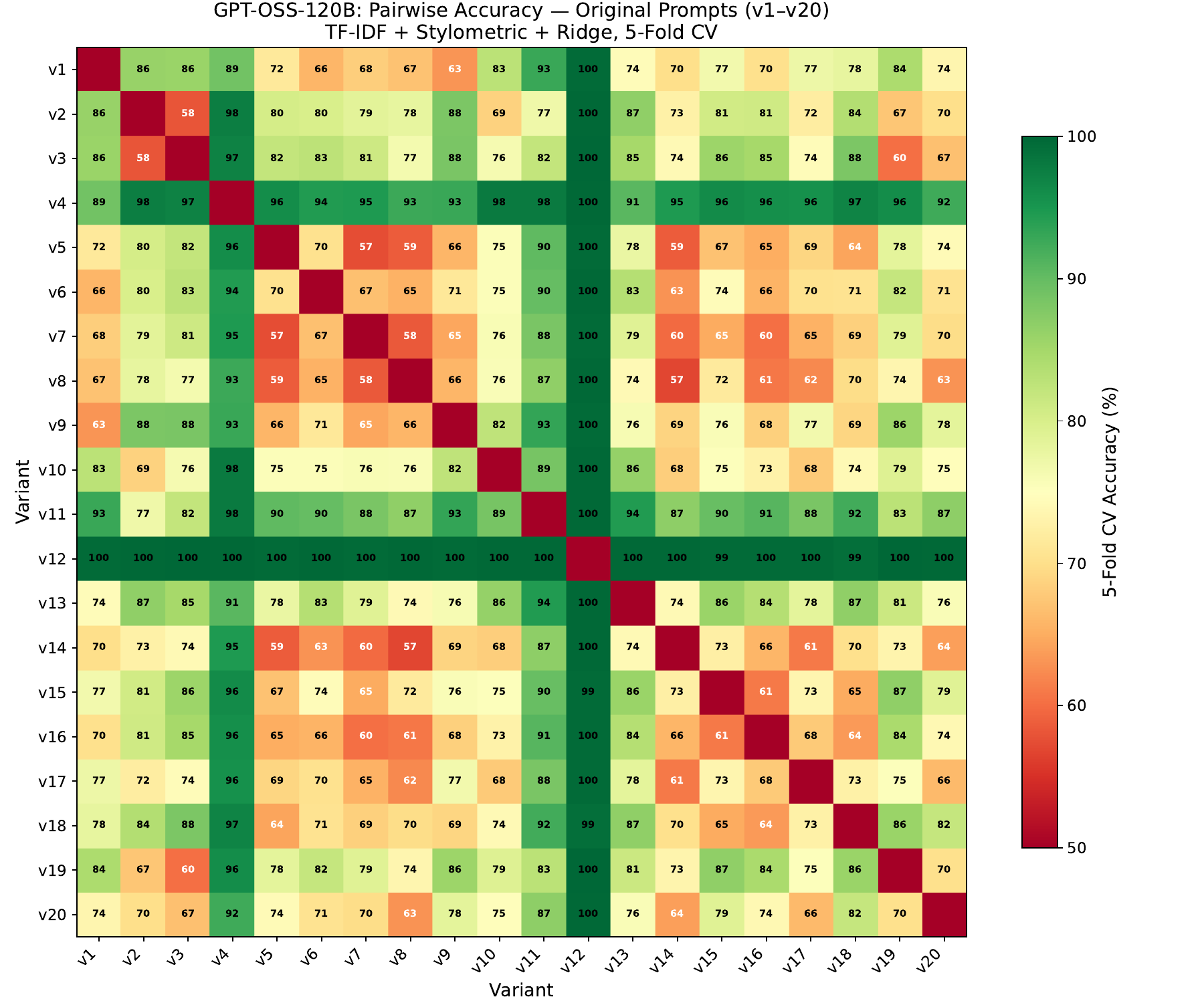}
    \caption{\textbf{\gptossonetwenty.}}
    \label{fig:gpt_oss_onetwenty_heatmap}
\end{figure}

\end{document}